\documentclass{vgtc}                          

\ifpdf
  \pdfoutput=1\relax                   
  \usepackage{graphicx}                
  \DeclareGraphicsExtensions{.pdf,.png,.jpg,.jpeg} 
\else
  \usepackage{graphicx}                
  \DeclareGraphicsExtensions{.eps}     
\fi%

\graphicspath{{figures/}{pictures/}{images/}{./}} 

\usepackage{microtype}                 
\PassOptionsToPackage{warn}{textcomp}  
\usepackage{textcomp}                  
\usepackage{mathptmx}                  
\usepackage{times}                     
\usepackage{cite}                      
\usepackage{tabu}                      
\usepackage{booktabs}                  
\usepackage{mathptmx}
\usepackage{epstopdf}

\usepackage{times}
\usepackage[lined,algonl,boxed]{algorithm2e}
\usepackage{amsthm, amssymb, amsmath}
\usepackage{mathrsfs}
\usepackage{dsfont}
\usepackage{multirow}

\usepackage{paralist}
\usepackage{tikz}
\usepackage{multirow}
\usepackage{color}
\usepackage{setspace}
\usepackage{microtype,hyphenat,balance}
\usepackage{array}
\usepackage{url}
\usepackage{bm}
\usepackage{algorithm2e}
\usepackage{algpseudocode}
\usepackage{stfloats}
\usepackage{float}
\usepackage{booktabs}
\usepackage{cite}
\newcommand{\jiazhi}[1]{\textcolor{black}{#1}}

\newcommand{\yang}[1]{\textcolor{black}{#1}}
\newcommand{\lei}[1]{\textcolor{black}{#1}}

\onlineid{1064}

\vgtccategory{Research}

\vgtcinsertpkg

\title{SMAP: A Joint Dimensionality Reduction Scheme for Secure Multi-Party Visualization}

\author{Jiazhi Xia\thanks{e-mail:xiajiazhi@csu.edu.cn}\\ %
    \parbox{1.4in}{\scriptsize \centering School of Computer Science and Engineering\\Central South University}
\and Tianxiang Chen\thanks{e-mail: txchen@csu.edu.cn}\\ %
     \parbox{1.4in}{\scriptsize \centering School of Computer Science and Engineering\\Central South University}
\and Lei Zhang\thanks{e-mail: mike\_lei@csu.edu.cn}\\ %
     \parbox{1.4in}{\scriptsize \centering School of Computer Science and Engineering\\Central South University}
\and Wei Chen\thanks{e-mail: chenwei@cad.zju.edu.cn}\\ %
     \parbox{1.4in}{\scriptsize \centering State Key Lab of CAD\&CG\\ Zhejiang University} 
\and Yang Chen\thanks{e-mail: chen1984yang@gmail.com}\\ %
     \parbox{1.4in}{\scriptsize \centering I4 data}
\and Xiaolong Zhang\thanks{e-mail: lzhang@ist.pus.edu}\\ %
     \parbox{1.4in}{\scriptsize \centering Pennsylvania State University, University Park, PA}
\and Cong Xie\thanks{e-mail: coxie@cs.stonybrook.edu}\\ %
     \parbox{1.4in}{\scriptsize \centering  Facebook}
\and Tobias Schreck\thanks{e-mail: tobias.schreck@cgv.tugraz.at \newline Wei Chen is the corresponding author.} \\
     \parbox{1.4in}{\scriptsize \centering Graz University of Technology}
}

\teaser{
  \centering
  \subcaptionbox*{(a)\vspace{3pt}}{
		 \includegraphics[width=0.32\columnwidth]{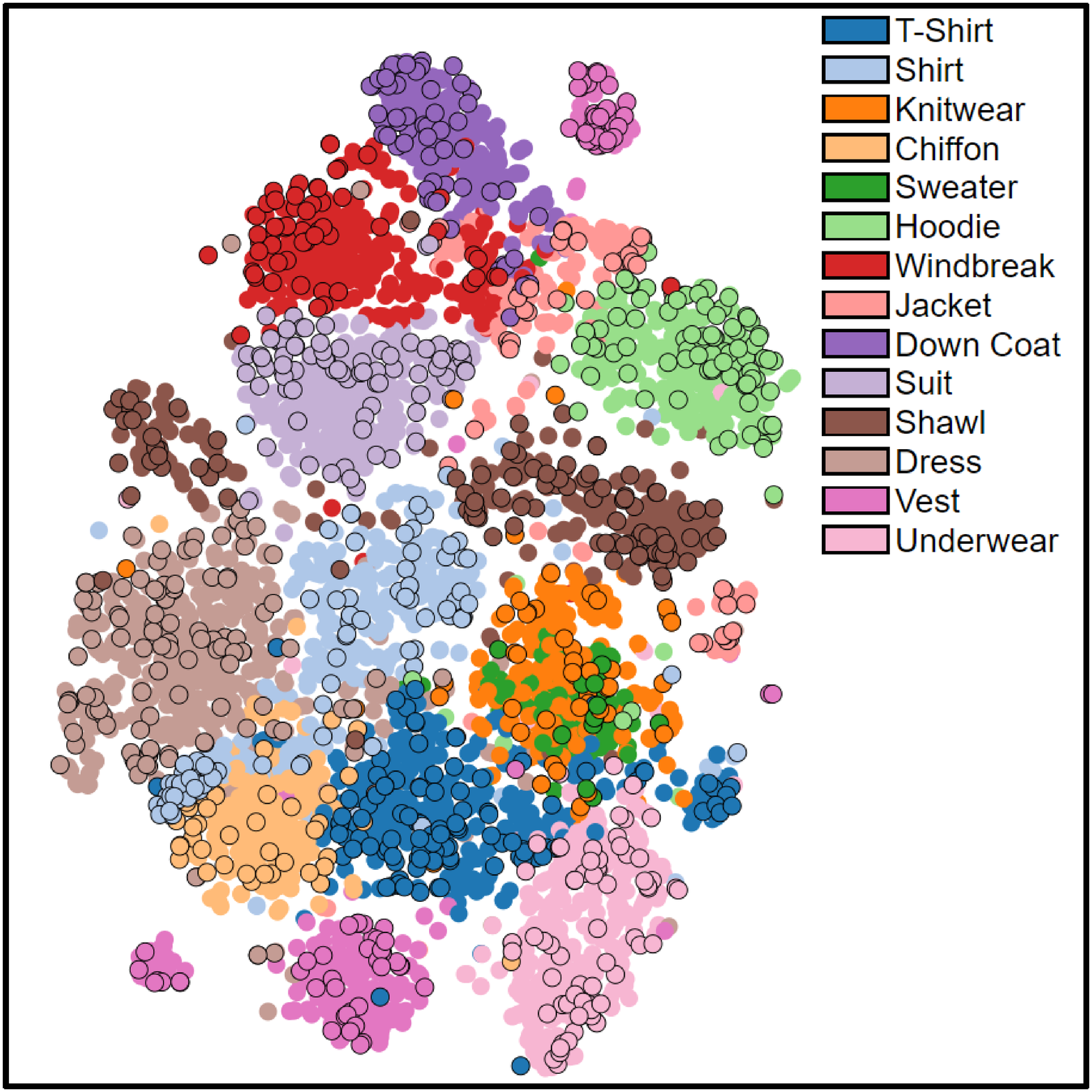}
	}
    \subcaptionbox*{(b)\vspace{3pt}}{
         \hspace{-2mm}
		 \includegraphics[width=0.32\columnwidth]{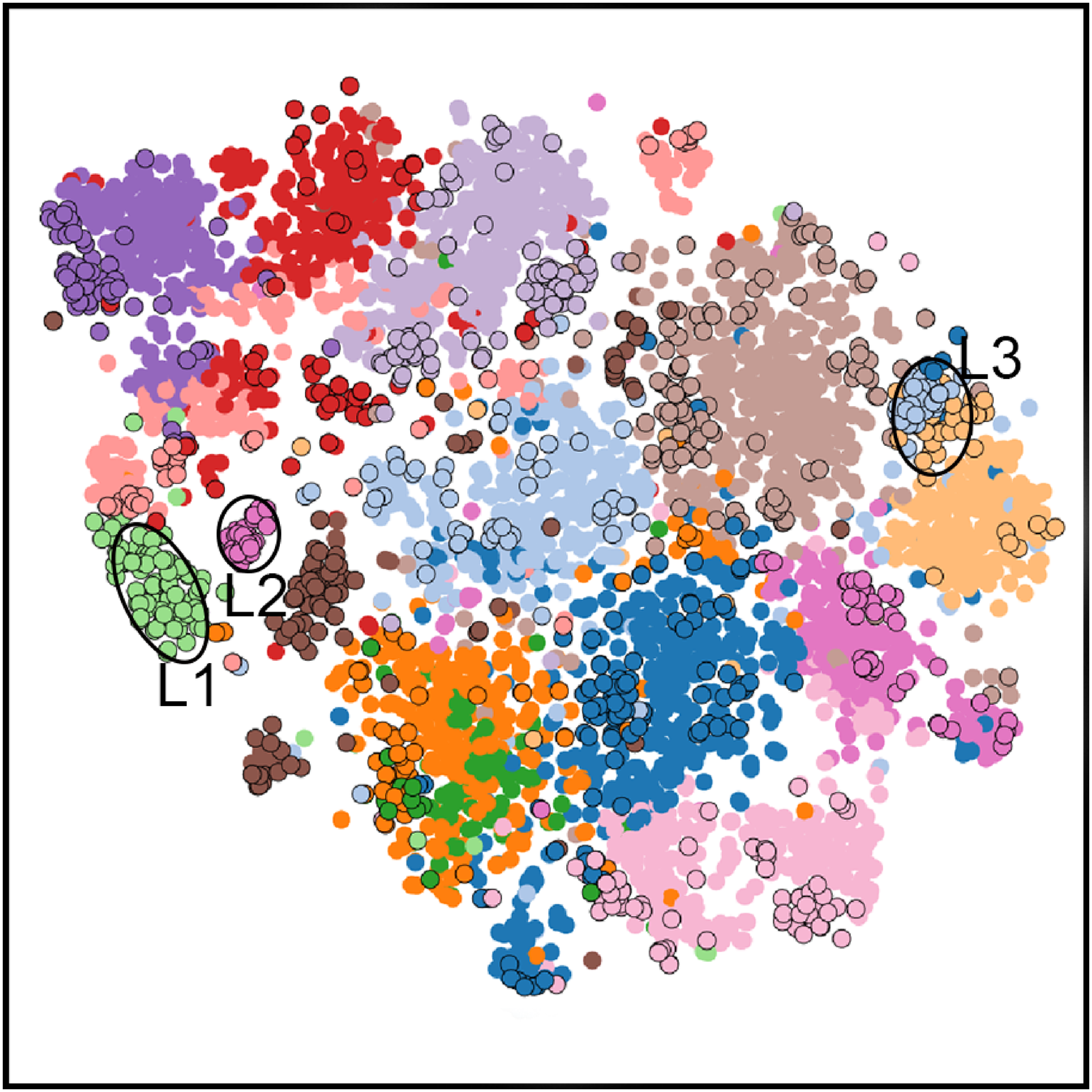}
	}
    \subcaptionbox*{(c)\vspace{3pt}}{
         \hspace{-2mm}
		 \includegraphics[width=0.32\columnwidth]{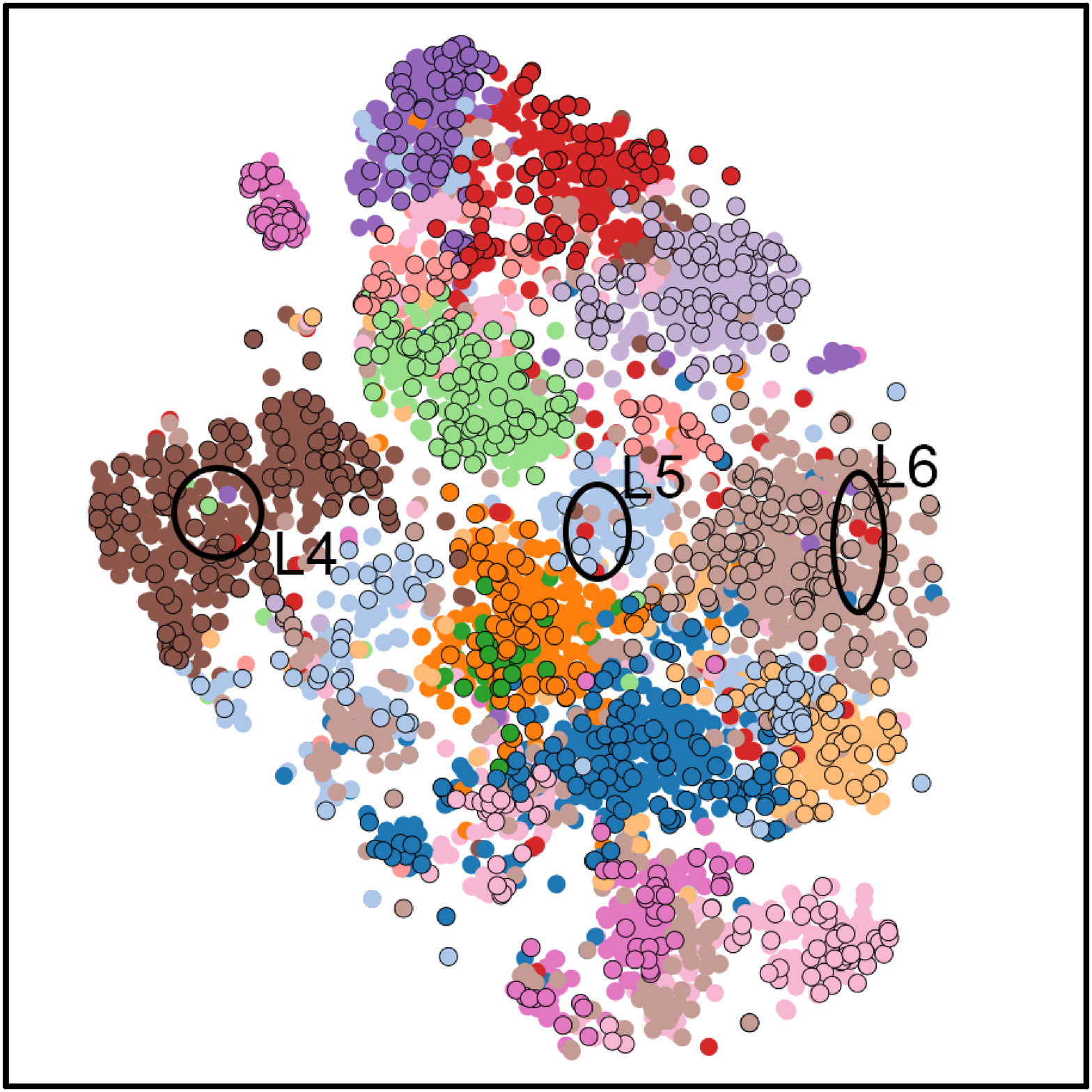}
	}
    \caption{A user performs three joint t-SNE projections with three data providers to compare the data quality. (a)--(c), the joint t-SNE results with dataset A, B, and C, respectively. The class labels are encoded by colors. Users' local data points are denoted by black borders.}
	\label{fig:teaser}
}

\abstract{Nowadays, as data becomes increasingly complex and distributed, data analyses often involve several related datasets that are stored on different servers and probably owned by different stakeholders. While there is an emerging need to provide these stakeholders with a full picture of their data under a global context, conventional visual analytical methods, such as dimensionality reduction, could expose data privacy when multi-party datasets are fused into a single site to build point-level relationships. In this paper, we reformulate the conventional t-SNE method from the single-site mode into a secure distributed infrastructure. We present a secure multi-party scheme for joint t-SNE computation, which can minimize the risk of data leakage. Aggregated visualization can be optionally employed to hide disclosure of point-level relationships. We build a prototype system based on our method, SMAP, to support the organization, computation, and exploration of secure joint embedding. We demonstrate the effectiveness of our approach with three case studies, one of which is based on the deployment of our system in real-world applications.
} 

\CCScatlist{
  \CCScatTwelve{High-dimensional data visualization}{secure visualization}{dimensionality reduction}{secure multi-party  computation};

}

\usepackage[font=scriptsize]{subcaption}

\begin{document}


\firstsection{Introduction}

\maketitle
Nowadays, as data is often distributed at multiple privacy-sensitive sites, individual data owners want to overcome this ``isolated islands'' problem~\cite{FederatedLearning2019Yang} to gain a full picture of their data with a global context and without leaking their own local data. Such a need may come from several hospitals that want to perform a joint analysis of their patient data while keeping their data privacy, or from a buyer in a data market who has to check the quality of several datasets from commercial data providers before making a data-purchasing decision~\cite{marketplace}.

When using visualization techniques to help people understand multi-party high-dimensional datasets, we often employ dimensionality reduction techniques, such as Principal Component Analysis (PCA)~\cite{wold1987pca},  Multidimensional Scaling (MDS)~\cite{kruskal1978MDS}, and t-distributed stochastic neighbor embedding (t-SNE)~\cite{maaten2008tsne}, to project all the data points into a shared space. However, this approach faces two challenges when applied to privacy-sensitive multi-party data analysis. First, data privacy can be breached when dimensionality reduction techniques are used to generate data for visualization. For example, conventional dimensionality reduction methods are designed for single-site computation and need to gather original data from all sites to create a joint embedding. Apparently, this approach can cause privacy leaks when original data is transmitted to other parties. 
\jiazhi{Simply anonymizing data before the multi-party dimensionality is also infeasible. 
Anonymized data would reduce the utility of data and lead to inaccurate layouts. In addition, even anonymized data possesses values that need to be protected.}
Second, data privacy may be threatened when other people use certain visualization designs to examine privacy-sensitive data. In particular, when visualization users are also data providers, they may have the authority to explore the low-dimensional embedding of data from other parties. The similarity in the low-dimensional space could disclose point-level relationships to some extent.

To address these challenges, secure multi-party visualization is needed. We define ``secure multi-party visualization'' as a method to visualize datasets from multiple parties in a shared space while keeping data privacy of each party.
In this paper, we propose a joint dimensionality reduction scheme for secure multi-party visualization. Our scheme can help to preserve data privacy at both visualization generation stage and visualization consuming stage. First, we develop a secure multi-party computation scheme for t-SNE embedding.
\jiazhi{We choose t-SNE because it is one of the most popular 
dimensionality reduction methods in visualization. In addition, it is reported to achieve high overall quality results in common tasks including cluster separation~\cite{espadoto2019towards}. The essential challenge in redesigning t-SNE is that both the original data and the distance matrix among data points can threaten data privacy.
Neither of them should be exposed to any participant. Therefore, we introduce two collaborators into the scheme. One collaborator can have only encrypted data and distance matrix, and the other can only access the noised data and distance matrix. Most importantly, the accurate t-SNE embedding can be computed with the noised distance matrix in our design. Both the data and distance matrix are kept private from participants.}
Second, we build an online visualization system, SMAP (Secure Multi-pArty Projection), to organize the joint embedding tasks among multiple participants and support privacy-aware exploration of embedding results.
In particular, our method supports two levels of visualization authority.
Scatterplots allow participants to have point-level embedding information of all datasets.
Aggregated visualization that is based on density map supports stricter requirements, with which the point-level relationships are hidden from other participants.
In addition, a set of descriptive views are coordinated to support task organization and data exploration.

The major contributions of this paper include:
\begin{compactitem}
  \item a secure multi-party dimensionality reduction scheme for joint t-SNE; and
  \item a coordinated online visualization system to support privacy-aware exploration and online task organization.
\end{compactitem}

In the rest of the paper, we review related work in Section~\ref{sec:related} and give an overview of our approach in Section~\ref{sec:overview}.
We describe the joint embedding in Section~\ref{sec:projection} and the online visualization system in Section~\ref{sec:vis}.
Then, we demonstrate case studies in Section~\ref{sec:case} and discuss our approach in Section~\ref{sec:discuss}.
The paper concludes in Section~\ref{sec:conclusion}.
\section{Related Work}
\label{sec:related}

\subsection{Security-aware Distributed Machine Learning and Data Mining}
\jiazhi{When raw data is distributed in multiple participants, it is challenging to build machine learning or data mining models among them due to security issues.}
A common approach to address the issue is applying the differential privacy theory \cite{dwork2006calibrating} that changes the raw input data with a randomized mechanism and keeps the utility of data. In practice, random perturbation, such as via Gaussian noise \cite{abadi2016deep} and Laplacian noise \cite{melis2015efficient}, is often used. 
However, while adding more noise can preserve privacy, this approach may compromise accuracy significantly and lead to interpretation bias of resulting patterns.

~\jiazhi{Many research efforts have been devoted to Secure Multi-Party Computation (SMPC) for distributed machine learning and data mining applications. Based on cryptographic protocols, the raw data are computed without perturbation in a secure manner. The SMPC schemes are
built upon classical cryptographic protocols~\cite{41}, including 
garbled circuits~\cite{SecureYao86, 16}, homomorphic encryption~\cite{addtive99Pascal,fully09Craig}, oblivious transfer~\cite{45,46}, and secret sharing~\cite{68,69}.  To support secure outsourced computation of k-means clustering, Almutairi et al.~\cite{almutairi2017k} designed a protocol based on fully homomorphic encryption. However, their distance matrix among data points is exposed to the third party.
Almutairi et al.~\cite{almutairi2018data} proposed the secure chain distance matrix to avoid exposure. However, it contains order information only and cannot support dimensionality reduction approaches. To speed up the SMPC, Peter et al.~\cite{peter2013efficiently} designed a protocol based on additive homomorphic encryption (AHE) rather than fully homomorphic encryption. Their protocol contains two collaborators to reduce the communication between participants and collaborators. However, the issue of exposing the distance matrix remains to be addressed.}


\subsection{Privacy-Preserving Visualization}
In the visualization domain, a variety of data types often involve sensitive information, such as health records for multi-dimensional data~\cite{dasgupta2014opportunities} or events for EMRs data~\cite{chou2019privacy}. Imposing privacy-preserving operations to visualization can prevent unauthorized disclosure of sensitive information while  visualization utility can be maintained and user insight gained from visualization can be maximized. Similar to the counterpart approaches in the data mining field \cite{agrawal2001design}, the protection aims at two major risk types: identity disclosure and attribute disclosure. While the former prevents the individual identity from being linked to a specific data item, the latter protects sensitive attributes that can be inferred from linking individuals to other attributes. A common strategy to address the risks is to apply visual uncertainty (e.g., blur effects or aggregated clusters) in the display of the sensitive information \cite{chou2019privacy}. For example, Dasgupta et. al. \cite{dasgupta2011adaptive} developed an enhanced parallel coordinates to visualize sensitive multi-dimensional data where private records are displayed as visual clusters between pairwise axes. To further assess the trade-off of privacy protection against visualization utility, a set of quantification measures have been proposed \cite{dasgupta2013measuring}. Wang et. al. \cite{wang2017utility} further improved the approach by integrating a pipeline to support interactive control over visualization utility and the level of privacy protection. Other privacy-preserving operations, such as redirecting links for graph layouts or directly deleting anomaly patterns containing sensitive information, are also discussed in \cite{chou2019privacy}.

In contrast to addressing the privacy issues in the original, high-dimensional visual space, our approach aims at embedding it in the low-dimensional space upon which the disclosure of identity and attribute can be naturally eliminated while substantial patterns such as clusters or anomalies can be preserved. If the point-level relationships are highly sensitive in stricter cases, aggregated visualization can be employed to hide the point-level information.

\subsection{Dimensionality Reduction}
\jiazhi{Dimensionality reduction techniques are well studied for scatterplot-based high-dimensional data visualization. Intuitively, they can project the data into low-dimensional space while preserving specific properties. For example, Principal Component Analysis (PCA)~\cite{wold1987pca} preserves the variance among data points. Linear Discriminant Analysis (LDA)~\cite{belhumeur1997lda} tries to keep clusters separable from each other. The t-distributed Stochastic Neighbor Embedding (t-SNE)~\cite{maaten2008tsne} and various variances of Multidimensional Scaling (MDS)~\cite{kruskal1978MDS} are designed to preserve the similarities among data points based on the distance matrix. Similarly, ISOMAP~\cite{tenenbaum2000isomap} aims to keep the geometric distance in the projected space. Recently, UMAP~\cite{mcinnes2018UMAP} and BH-tSNE~\cite{pezzotti2016hierarchical} require to compute the distances among the k-nearest neighbors. More details of dimensionality reduction methods can be found in surveys~\cite{liu2016visualizing,espadoto2019towards}.
}

\jiazhi{Because dimensionality reduction compresses data information, it is used to preserve data privacy in a wide range of applications. A set of dimensionality reduction methods are designed to discard the privacy-sensitive information and simultaneously preserve the utility of data in specific applications,such as machine learning~\cite{nguyen2020autogan}, data mining~\cite{liu2006randomproj}, and distance-based classification~\cite{Alotaibi2012DR}. It 
should be noted that the motivation of these methods is different from ours. They utilize dimensionality reduction to address privacy issue when publishing it for further applications. The dimensionality reduction is performed in a single site. In contrast, we aim to prevent data exposure from other participants during the multi-party dimensionality reduction. The dimensionality reduction is conducted among multiple participants who are privacy-sensitive.}

\jiazhi{The work by Saha et al.~\cite{debbrata17ijcai} is most similar to ours.
This work proposed to build a joint t-SNE among multiple participants. While pairwise distances were unavailable due to security issues, their method built embeddings around a shared dataset locally and iteratively integrated the embeddings as one. However, the requirement of the shared dataset limits the application in many private data analysis scenarios. In addition, the quality of the joint embedding depends on the volume of the shared dataset. As reported in Saha et al.~\cite{debbrata17ijcai}, heavy overlaps among clusters would happen with insufficient shared data. By contrast, we propose a Security Multi-party Computation framework for accurate t-SNE layout generation without requiring shared dataset.}

\section{Scenarios, Requirements, and Approach Overview}
\label{sec:overview}

In this section, we present two use scenarios to clarify the motivation and application of secure multi-party visualization.
Through the analysis of the two scenarios, we conclude the requirements of participants in a secure multi-party visualization.
Then, we overview the proposed approach.

\subsection{Example Scenarios}

\textbf{Multi-party data analytics.}
In this scenario, a group of participants want to analyze local data in the context of all global data when keeping data privacy \jiazhi{from each other}.
For example, analysts from multiple community hospitals want to explore differences among communities and identify abnormal patients.
However, they cannot fuse their data due to privacy concerns.
With the secure joint embedding, they can project local data points in a shared space, where the similarity among all data points can be visualized and analyzed.
The global patterns and distributions of each community are disclosed in the shared visualization.
With the full picture, these hospitals can visually locate their own records in the global context.
For instance, a hospital can verify if there is any similar patient to a special local case in other hospitals.
These insights can also guide further communications among the participating hospitals for further information rather than just low-dimensional layout.
If there are stricter constraints on data privacy that prohibits the disclosure of point-level relationships, these hospitals could only share the aggregated visualization for distribution information.

\textbf{Data purchasing.}
In data-driven applications, users often need to buy rich data to train their machine learning model through data markets~\cite{dataprice19} (e.g., BDEX~\cite{bedx} and Qlik~\cite{Qlik}).
Considering the high price of datasets, users must carefully select appropriate datasets, for example, that covers the failed samples or supports the rare classes.
However, there is no scheme that allows users to know the structure of a dataset before they buy it~\cite{marketplace, dataprice19}.
Secure multi-party visualization provides a tool for users to compare their local data with data on sale without data privacy being scarified.
This information transparency helps users to make purchase decisions.
From the perspective of commercial data providers, secure multi-party visualization allows them to demonstrate their data to users without losing their data.
This tool can promote data selling.
In this scenario, data providers can decide the strategy of visualization, such as scatterplots with point-level information, or aggregated visualization with only region-level information.

\subsection{Requirement Analysis}
Through the analysis of these application scenarios, we conclude the major requirements on a secure multi-party visualization scheme.

\textbf{R1: Preserve the privacy during the multi-party visualization generation.}
Security is the major concern of the participants.
Both the constraints of General Data Protection Regulation (GDPR) and the need to protect data assets prevent participants from revealing the original data to others.
Considering the fact that all participants are curious about data, it should be guaranteed that all participants cannot obtain or infer the original data in the entire process.

\textbf{R2: Preserve the privacy during the visualization consuming.}
The second difference between secure joint embedding and conventional joint embedding is that privacy should also be preserved at the visualization consuming stage.
In a scatterplot, each participant should only be allowed to explore the low-dimensional representation of global data points rather than high-dimensional representations.
If point-level relationships are also considered, aggregated visualization should be employed rather than scatterplots.

\textbf{R3: Support interactive exploration of privacy-aware visualization results.}
With the visualization results, participants need to know the structure of global data, compare the distributions of participants, and analyze the high-dimensional information of interested data from other participants.
The difference between the exploration of secure joint embedding and conventional embedding results is that only limited information is available in secure joint embedding because of privacy concerns.
While only low-dimensional layout of global data is provided to each participant, they can only infer the high-dimensional information by local data and the similarities among local data and global data.
A set of coordinated views and interactive tools are needed to support privacy-aware exploration.

\textbf{R4: Support the organization of online joint embedding tasks among multiple participants.}
In addition to joint embedding algorithms, an online system to connect participants in different sites is needed to set the framework in a real environment.
The system should support participants to browse published tasks, propose tasks, join proposed tasks, and trace the participating tasks.
Because the process of multi-party visualization is time-consuming, it is important to keep following up on the progress and results of joint embedding tasks.

\subsection{Approach Overview}
Based on the above requirements, we propose a joint t-SNE scheme and develop an online visualization system for secure multi-party visualization.
First, we propose a secure multi-party protocol for joint t-SNE.
The protocol is designed to preserve data privacy.
Only encrypted and noised data are transferred to two external collaborators.
No explicit high-dimensional representations or the pairwise distances are exposed to any participants or collaborators.
The security of data is guaranteed with two non-collusion collaborators.
Second, we design an online visualization system that combines a set of coordinated views.
Privacy-preserving visualizations and privacy-aware explorations are supported by the system.
In addition, the online system connects the participants from different sites and organizes online joint embedding tasks.

\section{Secure Multi-Party Projection}
\label{sec:projection}

In this section, we describe our secure multi-party projection scheme for t-SNE.
It embeds distributed data points from multiple participants into a shared layout and preserves the privacy of local data.
We first briefly review the t-SNE algorithm, and then describe the protocol of our secure multi-party t-SNE.
After that, we analyze the security and accuracy of the proposed protocol.

\subsection{The t-SNE algorithm}

In high-dimensional space, the similarity between two data points $x_i$ and $x_j$ is represented by a symmetric probability $p_{ij}$, which means the probability that $x_i$ is a neighbor of $x_j$. $p_{ij}$ is given by

\begin{equation}
\label{eq:symhp}
p_{ij} = \frac{(p_{j|i}+p_{i|j})}{2N},
\end{equation}
where $N$ is the number of points, and the conditional probability $p_{j|i}$ is given by

\begin{equation}
\label{eq:hcp}
p_{j|i} = \frac{exp(-||x_i-x_j||^2/{2\sigma_i^2})}{\sum_{k\not=l}exp(-||x_i-x_k||^2/{2\sigma_i^2})},
\end{equation}
where $\sigma_i^2$ is the variance of the Gaussian that is centered on data point $x_i$, $||x_i-x_j||$ is the Euclidean distance between $x_i$ and $x_j$.

In low-dimensional space, t-SNE employs a t-distribution with one degree of freedom to describe the distribution.
The joint probability $q_{ij}$, which measures the similarity between points $y_i$ and $y_j$, is defined as

\begin{equation}
\label{eq:lcp}
q_{ij} = \frac{(1+||y_i-y_j||^2)^{-1}}{\sum_{k\not=l}(1+||y_k-y_l||^2)^{-1}},
\end{equation}
where $y_i$ and $y_j$ are the embedding of $x_i$ and $x_j$, respectively.

The t-SNE algorithm optimizes the embedding by minimizing the Kullback-Leibler divergence between the probability distributions in high-dimensional space and low-dimensional space. The Kullback-Leibler divergence is given by

\begin{equation}
\label{eq:KL}
C = KL(P||Q) = \sum_{i}\sum_{j}p_{ij}log\frac{p_{ij}}{q_{ij}}.
\end{equation}

\subsection{Secure Multi-Party Protocol for joint t-SNE}
Through the review of t-SNE algorithm, we know that the key step is to compute the conditional probability $p_{j|i}$ in the high-dimensional space.
Equation~\ref{eq:hcp} shows that the computation of $p_{j|i}$ depends on the computation of pairwise distances in high-dimensional space, which is a barrier of reformulating conventional dimensionality reduction approaches into a secure multi-participant version,
\jiazhi{because the pairwise distance matrix cannot be exposed to any participant.}
Therefore, the computation of $p_{j|i}$ must be redesigned.

\begin{figure}[!tb]
	\centering
\includegraphics[width=\columnwidth]{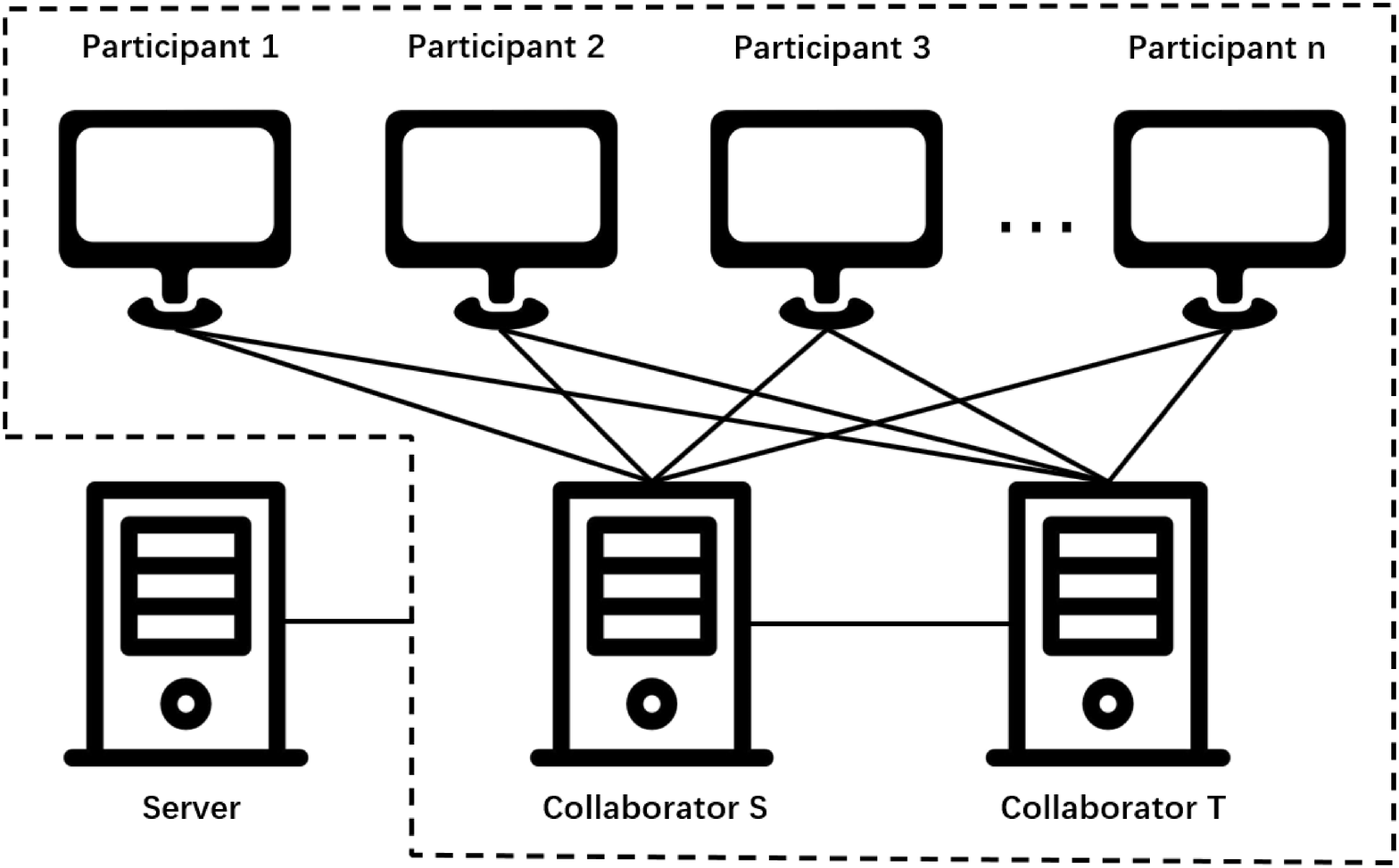}
\caption{The architecture of SMAP. The system is set in a server to organize online joint embedding tasks. The server connects and organizes the two collaborators and multiple participants. Collaborators $S$ and $T$ are distributed on two different sites and connected to each other. The participants are connected to the two collaborators.}
	\label{fig:architecture}
\end{figure}

\jiazhi{To address this issue, we propose a secure multi-party protocol for t-SNE.
In addition to the participants ($P$) who own data, two collaborators, $S$ and $T$,
are introduced into the protocol to provide computing service. The architecture is presented in Figure~\ref{fig:architecture}. Due to privacy concerns, $T$ has only encrypted data without the private key, while $S$has only noised data.}
$S$ and $T$ cooperate with each other to compute the t-SNE layout and distribute the results to the participants.
Data privacy is preserved under the assumption that these two collaborators will not collude with each other.
All the participants are semi-honest and will execute the protocol correctly.
While credit-driven collaborators exist in reality, non-collusion and semi-honest models are practical for large-scale data processing~\cite{SMPC19zhao}.

\begin{figure}[!tb]
    \centering
    \includegraphics[width=1.0\linewidth]{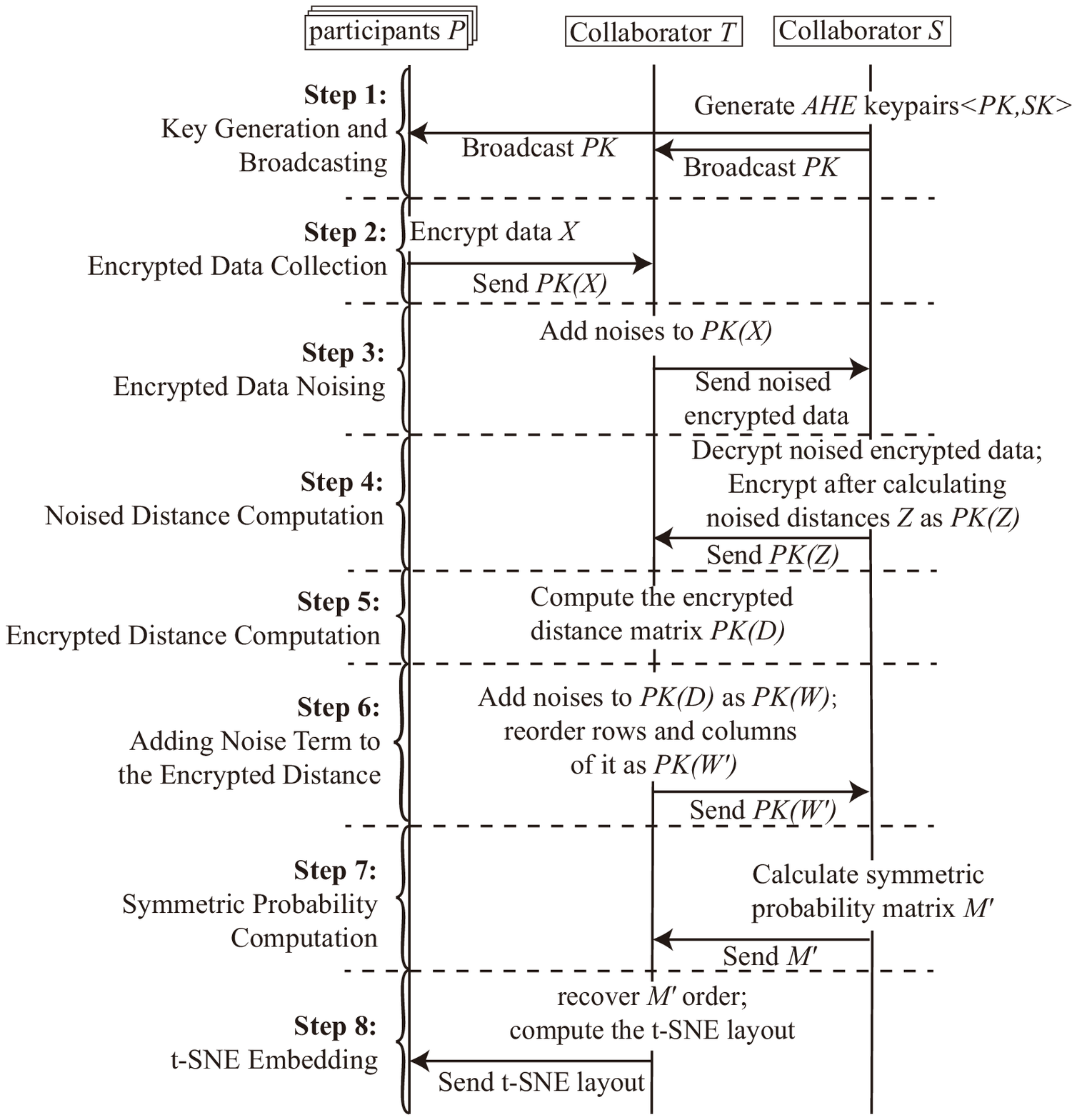}
    \caption{The pipeline of secure multi-party protocol for joint t-SNE.}
    \label{fig:algorithm}
\end{figure}

The process of our protocol is illustrated in Figure~\ref{fig:algorithm}.
For the clarity of description, we list the notations in Table~\ref{tab:notation}.
The protocol contains eight steps of data transmission and computation.

\begin{table}[!tb]
 \caption{\label{tab:notation} The notation definitions.}
 \begin{tabular}{lp{0.78\linewidth}}
\toprule
  Notation & Description \\
\midrule
  $P$ & Participating parties who are the date owners. \\
  $S$, $T$ & Two collaborators who perform the secure computing.\\
  $PK$ & The public key. $PK()$ is the encryption operation.\\
  $SK$ & The private key. $SK()$ is the decryption operation.\\
  $\circ, \diamond, \star$ & The addition, subtraction, and scalar multiplication operations for encrypted data, respectively.\\
  $x,X$ & $x$ is a high-dimensional data point. $X$ is the collection of high-dimensional data.\\
  $\overline{x_{ij}}$ & $\overline{x_{ij}}$ is the noised entry as $\overline{x_{ij}} = x_{ij}+\sigma_{ij}$.\\
  $\sigma_{ij}$ & The noise term of $\overline{x_{ij}}$.\\
  $\delta_{ijk}$ & A noise term. $\delta_{ijk} = \sigma_{ik}-\sigma_{jk}$.\\
  $d_{ij}, D$ & $d_{ij}$ is the Euclidean distance between $x_i$ and $x_j$. D is the corresponding distance matrix.\\
  $z_{ij}, Z$ & $z_{ij}$ is the squired noised distance between $x_i$ and $x_j$ regarding to noise term $\sigma_{ij}$. $Z$ is the corresponding squired noised distance Matrix between all data point pairs.\\
  $w_{ij}, W$ & $w_{ij}$ is the noised distance between $x_i$ and $x_j$, $w_{ij} = d_{ij} + \eta_i$. $W$ is the corresponding noised distance matrix.\\
  $\eta_i$ & The noise term of the $i$th row in $W$.\\
  $W'$ & The result of randomly reordering $W$ by row and by column.\\
  $M$ & The symmetric probability matrix.\\
  $M'$ & The reordered result of $M$, of which the indexes are the same as $W'$.\\
\bottomrule
\end{tabular}
\end{table}

\textbf{Step 1: Key generation and broadcasting.} $S$ generates the public key $PK$ for data encryption and the private key $SK$ for data decryption. $S$ broadcasts $PK$ to $T$ and all $P$. In our protocol, we used additive homomorphic encryption~\cite{addtive99Pascal} for data encryption. Under Additive Homomorphic Encryption (AHE), the addition and scalar multiplication are preserved. Specifically, we have $SK(PK(a) \circ PK(b))=a+b$, $SK(PK(a) \diamond PK(b))=a-b$, and $SK(a \star PK(b))=a*b$, where $PK()$ refers to encryption with the public key $PK$; $SK()$ refers to decryption using the private key $SK$, $\circ$, $\diamond$; and $\star$ are addition, subtraction, and scalar multiplication operations for encrypted data, respectively.

\textbf{Step 2: Encrypted data collection.} Each $P$ encrypts local data $X$ with the public key $PK$ and sends the encrypted data $PK(X)$ to $T$.

\textbf{Step 3: Encrypted data noising.}Given the data $X$ from all participating parties, $T$ adds a random noise to each entry of the encrypted data. Specifically, for each $x_{ij}$, which is the $j$th dimension of data point $x_i$, $T$ computes $PK(\overline{x_{ij}}) = PK(x_{ij}) \circ PK(\sigma_{ij})$, where $\sigma_{ij}$ is a random noise. Then, $T$ sends the noised encrypted data to $S$.

\textbf{Step 4: Noised distance computation.} First, Collaborate $S$ decrypts the noised data with $SK(PK(\overline{x_{ij}})) = x_{ij} + \sigma_{ij}$. Second, the noised Euclidean distance $z_{ij}$ between $x_i$ and $x_j$ is computed as

\begin{equation}
\label{eq:noised}
z_{ij}=\sum_{k=1}^m((x_{ik}+\sigma_{ik})-(x_{jk}+\sigma_{jk}))^2,
\end{equation}
where $m$ is the dimensionality of data points. Third, $S$ encrypts the noised distances as $PK(Z)$ and sends to $T$, where $Z$ is the encrypted noised distance matrix.

\textbf{Step 5: Encrypted distance computation.} In this step, Collaborate $T$ computes the encrypted accurate distance matrix $D$.

In Equation~\ref{eq:noised}, we denote $\sigma_{ik}-\sigma_{jk}$ as $\delta_{ijk}$. The Equation~\ref{eq:noised} is rewritten as

\begin{equation}
\label{eq:noised2}
z_{ij}=\sum_{k=1}^m(x_{ik}^2+x_{jk}^2+\delta_{ijk}^2-2x_{ik}x_{jk}+2x_{ik}\delta_{ijk}-2x_{jk}\delta_{ijk}).
\end{equation}

Furthermore, we denote the Euclidean distance between $x_i$ and $x_j$ as $d_{ij}$ in Equation~\ref{eq:noised2}. We have

\begin{equation}
\label{eq:noised3}
z_{ij}=d_{ij}^2+\sum_{k=1}^m{(\delta_{ijk}^2+2x_{ik}\delta_{ijk}-2x_{jk}\delta_{ijk})}.
\end{equation}

Therefore, the encrypted squared distance $PK(d_{ij}^2)$ can be computed by

\begin{equation}
\label{eq:dis}
PK(z_{ij})\diamond \sum_{k=1}^m{(PK(\delta_{ijk}^2) \circ PK(2\delta_{ijk})\star PK(x_{ik}) \diamond PK(2\delta_{ijk})\star PK(x_{jk}))}.
\end{equation}

Because Collaborate $T$ records the noise term $\delta_{ijk}$ in Step 3, the scalar multiplication $\star$ between $PK(x_{ik})$ and $\delta_{ijk}$ is allowed. We denote the encrypted distance matrix as $PK(D)$.

\textbf{Step 6: Adding noise term to the encrypted distance.} In this step, Collaborate $T$ adds noises to the encrypted distance matrix $PK(D)$ and sends the noised matrix to collaborate $S$. For each entry $PK(d_{ij}^2)$ in the $j$th line of $PK(D)$ except for $PK(d_{ii}^2)$, $T$ adds a random noise $\eta_i$ to it as $PK(w_{ij}) = PK(d_{ij}^2) \circ PK(\eta_i)$, where $w_{ij}$ denotes the noised entry $d_{ij}^2$. If the entry $d_{ii}^2$ adds the same noise, $S$ can obtain the noise $\eta_i$ by $SK(PK(d_{ii}^2\circ PK(\eta_i)))$, because the $d_{ii}^2$ is always zero. The noised matrix is denoted as $PK(W)$. After that, $T$ reorders the rows and columns of $PK(w)$ and denotes the reordered matrix as $PK(W')$. The mapping between $PK(W)$ and $PK(W')$
is saved in $T$. Finally, $T$ sends $PK(W')$ to $S$.

\textbf{Step 7: Symmetric Probability Computation.} In this step, $S$ computes the symmetric probability in the high-dimensional space. Given the noised distance matrix $W'$ by $SK(PK(W'))$, we have


\begin{equation}\begin{split}\label{eq:probh}
        \lei{p'_{j|i}}&\lei{=\frac{exp(-(d_{ij}^2+\eta_i)/{2\sigma_i^2})}{\sum_{k\not=l}exp(-(d_{ik}^2+\eta_i)/{2\sigma_i^2})}}\\
        &\lei{=\frac{exp(-d_{ij}^2/{2\sigma_i^2})exp(-\eta_i/{2\sigma_i^2})}{\sum_{k\not=l}exp(-d_{ik}^2/{2\sigma_i^2})exp(-\eta_i/{2\sigma_i^2})}}\\
        &\lei{=\frac{exp(-d_{ij}^2/{2\sigma_i^2})}{\sum_{k\not=l}exp(-d_{ik}^2/{2\sigma_i^2})}=p_{j|i}}
\end{split}\end{equation}

While $\eta_i$ can be removed from Equation~\ref{eq:probh} by a fraction reduction, we have $p'_{j|i} = p_{j|i}$ according to Equation~\ref{eq:hcp}. Therefore, $S$ can compute the conditional probability $p_{j|i}$ with the noised distance matrix $W'$.
$p_{i|j}$ can be computed similarly. After that, $p_{ij}$ can be computed by Equation~\ref{eq:symhp}. We denote the symmetric probability matrix as $M'$. Finally, $S$ sends $M'$ to $T$. It is worth noting that the indexes in $M'$ are not the same as that in $W$.

\textbf{Step 8: t-SNE Embedding.} Having the reordered probability matrix $M'$, $T$ recover its order as $M$ according to the mapping between between $PK(W)$ and $PK(W')$ in Step 6. After that, $T$ computes the t-SNE embedding by optimizing Equation~\ref{eq:KL}.

\subsection{Protocol Analysis}
\textbf{Security.} As analyzed above, there are two privacy-sensitive data items in the entire process: the original data and the distance matrix. Having the full distance matrix, one can easily recover the original data with a few geometric computations based on a few data points. Therefore, we should guarantee that no participants can get any one of the two data items.
We analyze the three kinds of roles: participants $P$, collaborator $S$, and collaborator $T$.
First, the participants $P$ have only their local data and the distributed embedding results.
It is highly impractical to recover the accurate pairwise distances from the t-SNE embedding.
Therefore, the information security is guaranteed.
Second, $S$ has the public key $PK$, private key $SK$ (see Step 1), the noised encrypted data $x_{ij} + \sigma_{ij}$ (see Step 3), and the noised and reordered distance matrix $W'$ (see Step 7). Because the noises are known only by $T$, $S$ can recover neither the original data nor the distance matrix. The noised distance matrix $W$ is randomly reordered as $W'$, so that $S$ does not have the correct index to recover the row-based noise term $\eta$. Otherwise, $S$ can build linear equations by subtracting symmetric entries of $W$ and solve it for $\eta$.
Third, $T$ has only the encrypted data (see Step 2) and the encrypted distance matrix (see Step 5). It cannot recover the privacy-sensitive data items without the private key.
As a result, the security of data can be guaranteed in our protocol.

This security analysis has not considered that low-dimensional embedding would expose point-level relationships, because it is exactly the object of t-SNE. Instead, we will analyze this issue in Section~\ref{sec:vis}.

\vspace{3mm}
\noindent
\textbf{Accuracy.} The embedding result is exactly the same as that from the standard t-SNE algorithm.
It is worth noting that although noises are added to the pairwise distance in Step 6, the noise can be removed implicitly in Equation~\ref{eq:probh} by a fraction reduction (Step 7).
In Step 8, $T$ obtains the exact symmetric probability matrix $M$, which is required by a t-SNE algorithm (see Equation~\ref{eq:KL}). The following optimization is the same as the standard t-SNE algorithm. Therefore, our protocol presents a joint exact t-SNE embedding.

\section{SMAP System}
\label{sec:vis}

We designed an interactive online system, SMAP, to organize the visualization tasks among multiple participants (\textbf{R4}) and support the exploration of privacy-aware visualization results (\textbf{R3}).
Aggregated visualization for point-level privacy-preserving is integrated (\textbf{R4}).
In this section, we introduce the architecture, aggregated visualization, and the user interface of SMAP.

\subsection{Architecture and Implementation}


Figure~\ref{fig:architecture} shows the architecture of SMAP.
There are three kinds of roles in the system.
The server organizes the schedule of each joint embedding task and provides a platform that connects the collaborators and participants.
It has only the information for task organization but not the transferred data among the participants and collaborators.
The collaborators provide the computing resources, such as cloud computing, for joint embedding.

$S$ and $T$ belong to two different reputable entities without collusion.
Organized by the server, $S$ and $T$ are connected with each other and send encrypted or noised data to each other.
The participants are connected to the two collaborators following the secure multi-party protocol.
$S$ broadcasts the public key to them.
They send the encrypted data to $T$ and receive visualization results from $T$.
An client interactive interface is installed in each participant for task management and exploration of visualization results.

SMAP is developed as a web-based system. Its front end is implemented with JavaScript and D3.js.
The back end of the server is implemented with python.
We use a Python-based library for the partially homomorphic encryption~\footnote{https://python-paillier.readthedocs.io/en/develop/} and develop a multi-thread implementation.
In our case studies, the clients are browsed in laptops with an Intel i7-8750H CPU, which supports 12 threads, 16GB CPU memory, and a GeForce GTX 1060 with 6GB GPU memory.
Each server has an Intel Xeon E5-2683 v4 CPU (32GB memory), which supports 32 threads.

\subsection{User Interface for Participants}
\begin{figure*}[!tb]
	\centering
    \includegraphics[width=\linewidth]{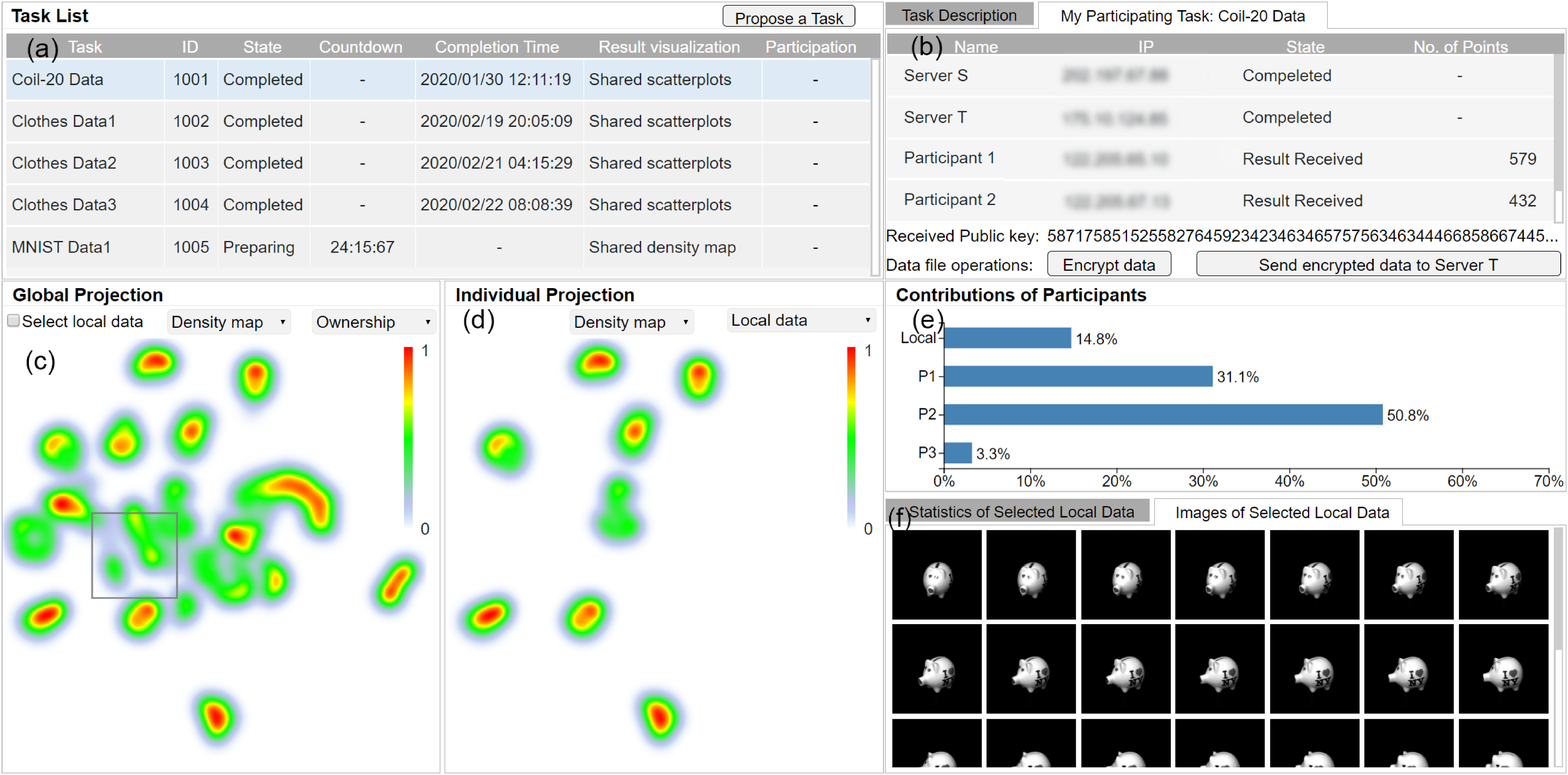}
    \caption{The user interface for participants. (a): the task list; (b): the task description view; (c): the global projection view; (d): the individual projection view; (e): the bar chart of contributions; and (f): the parallel coordinates/snapshot list.}
	\label{fig:interface}
\end{figure*}

In this section, we introduce the design and interaction of the privacy-aware user interface for participants (Figure~\ref{fig:interface}).
The interface contains the task organization panels (\textbf{R4}), visualization results of joint embedding (\textbf{R2}), and a set of descriptive views for the data.

\textbf{Online task organization.}
In the task list panel (Figure~\ref{fig:interface}a), participants can browse the online tasks, join in a preparing task, and propose a new task.
After selecting a task in the list, the task description panel (Figure~\ref{fig:interface}b).
 presents the task information, such as the contents of data and the participants.
For a participating task, participants can switch to the tab of "My Participating Task", which presents the details of all collaborators and participants, including their roles, IP addresses, states, and contributed data points.
This panel also supports the task operations, including data encryption and upload.

\textbf{Visualizations of joint embedding.}
The privacy-aware embedding results are presented in two views.
The global projection view (Figure~\ref{fig:interface}c) presents the embedding results of data from all participants.
In the individual projection view (Figure~\ref{fig:interface}d), only the data of the selected participant is displayed according to the global embedding.
To fulfill the privacy-preserving requirements, we provide two visualization strategies.
When participants want to know point-level relationships, the results are presented as scatterplots, in which color encodes data ownership or data class.
In the scatterplot mode, participants can freely select local points with a lasso tool in these two views.
Because participants are limited to access the original local data, they can only infer the high-dimensional representation of data of others by selecting nearby local data.

When participants are aware of point-level privacy, aggregated visualization is provided to show the joint embedding results.
Mostly popular designs for aggregated visualization include  binning~\cite{binning2018Heimerl} and density map~\cite{splatterplots2013Mayorga}.
Figure 5 compares the visualization results  between these two designs.

\begin{figure}[!tb]
	\centering
	\subcaptionbox{\vspace{5pt}}{
        \vspace{-3pt}
		\includegraphics[width=0.3\columnwidth]{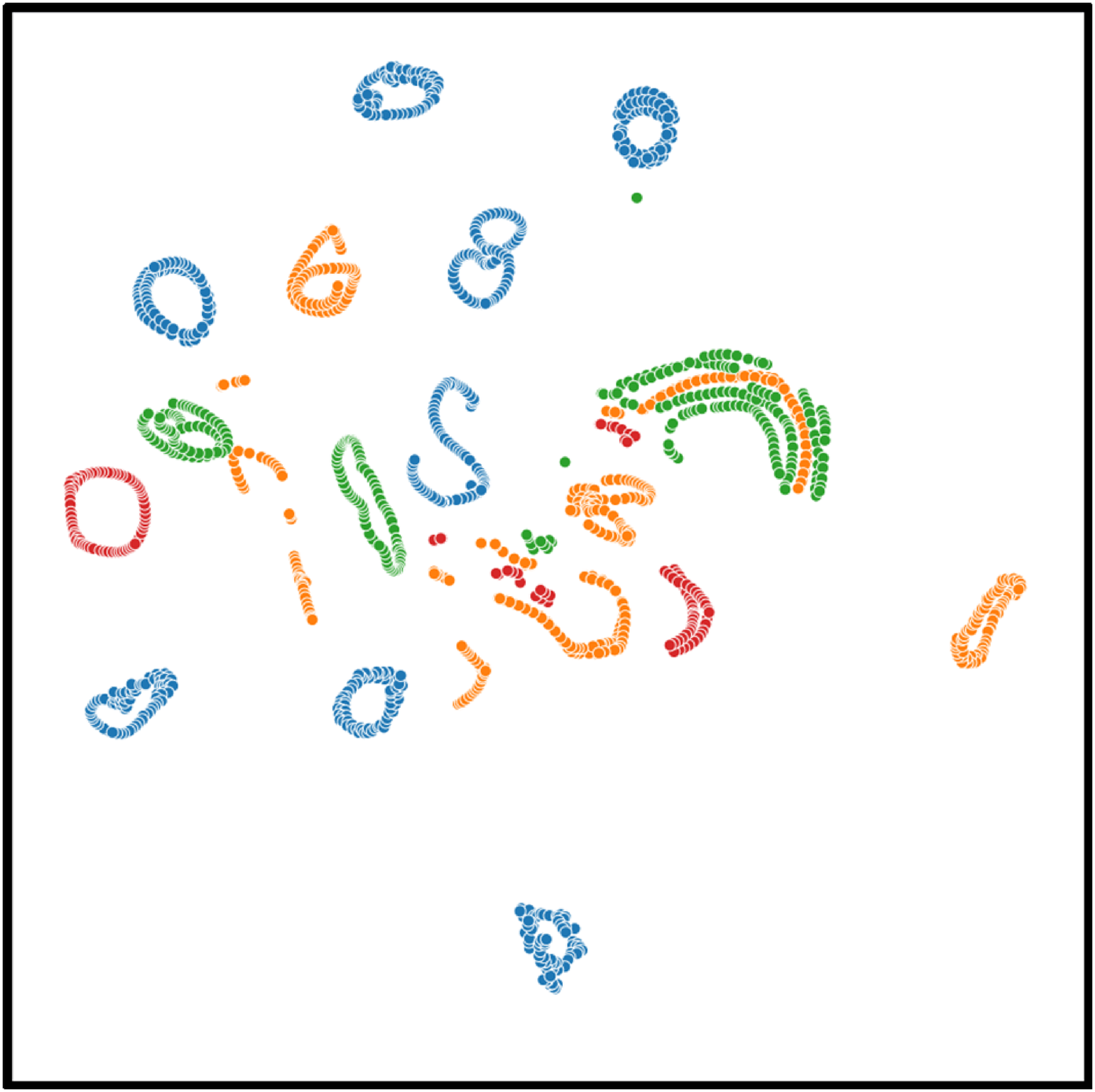}
	}	
		\subcaptionbox{\vspace{5pt}}{
        \vspace{-3pt}
	    \includegraphics[width=0.3\columnwidth]{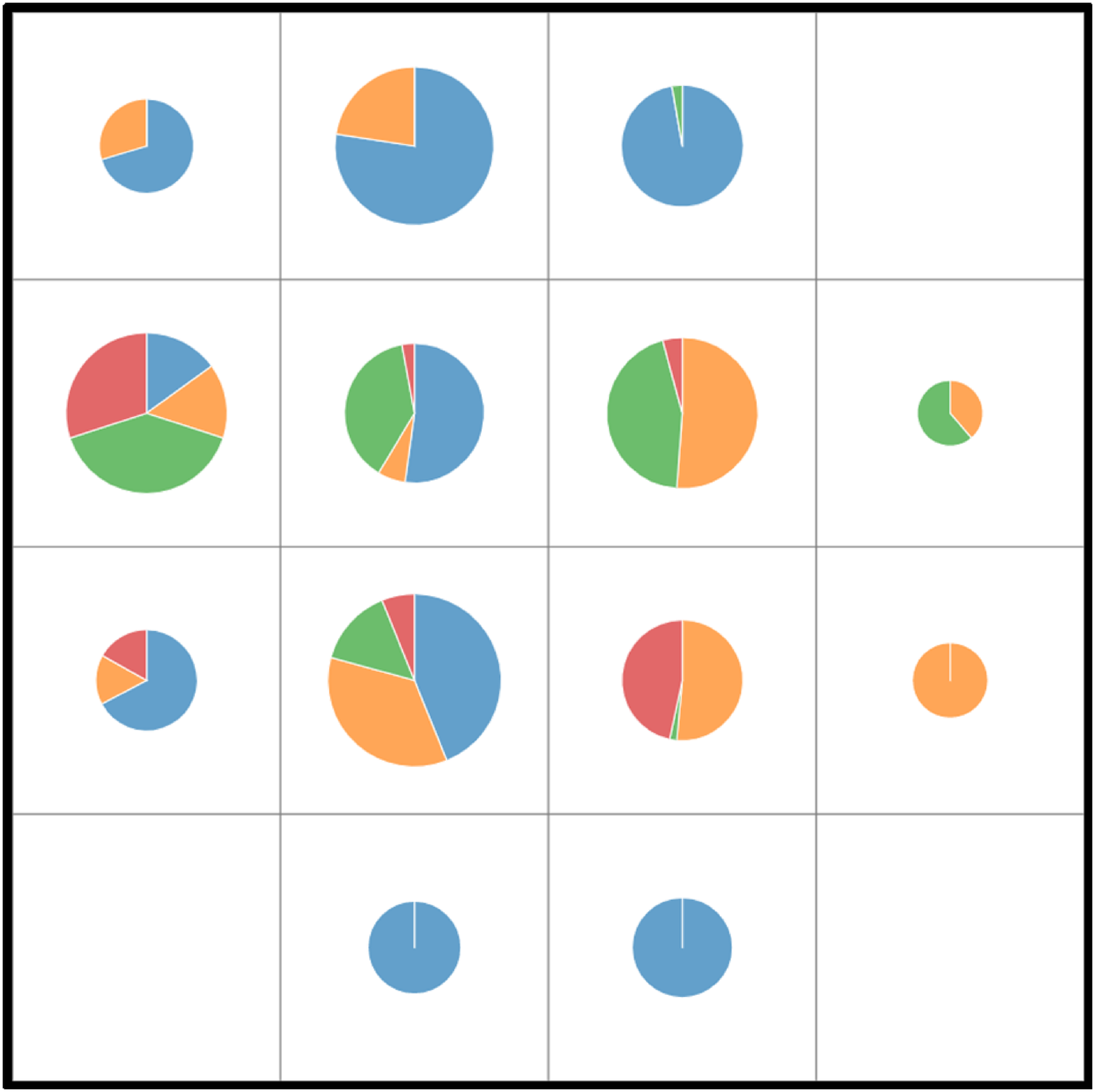}
	}
	\subcaptionbox{\vspace{5pt}}{
        \vspace{-3pt}
	    \includegraphics[width=0.3\columnwidth]{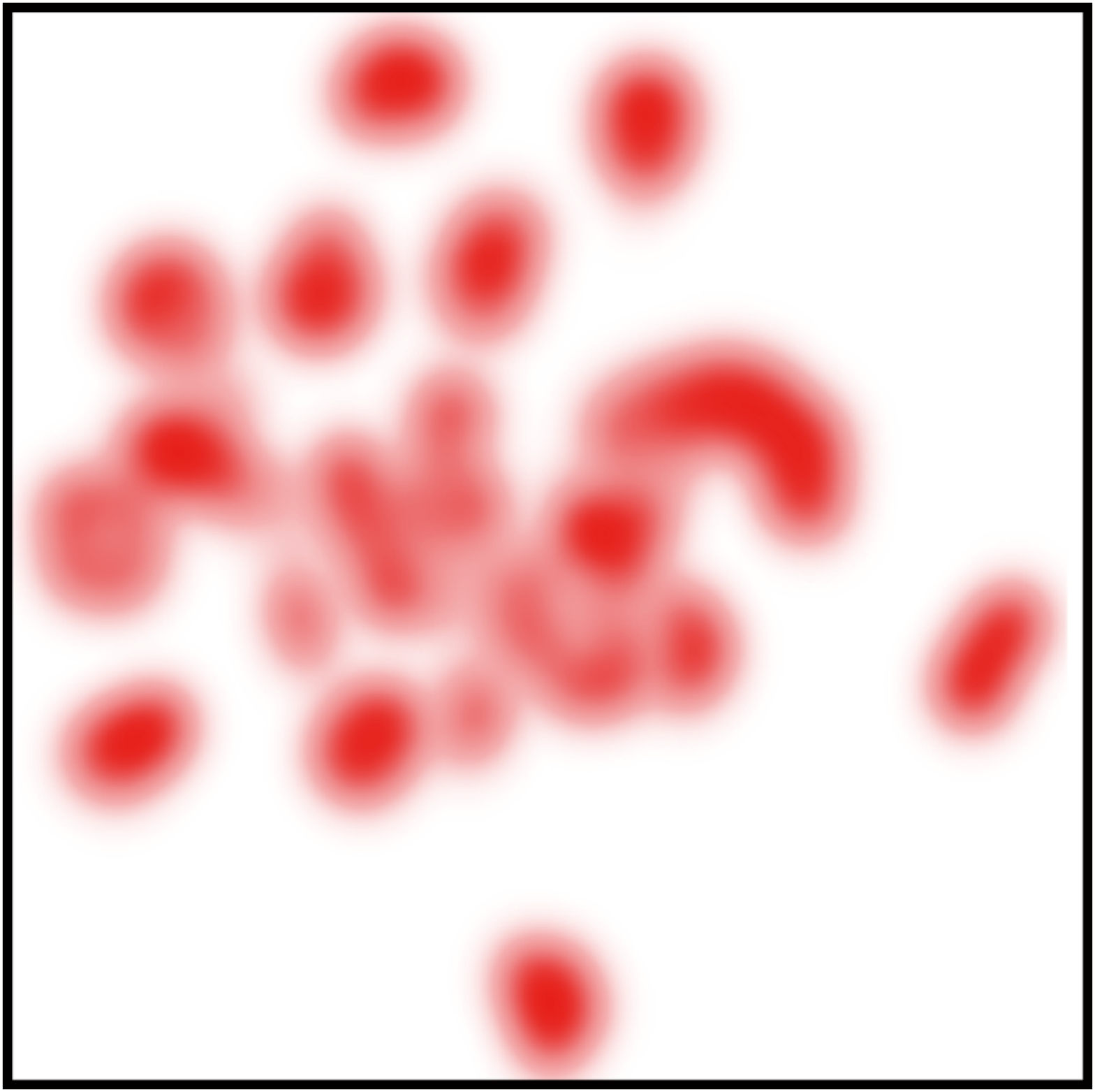}
	}
	\caption{Design choices for aggregated visualization. Color encodes data ownership in the scatterplots and binning plots. (a) the original result of joint t-SNE; (b) the binning design; (c) the density map. }
	\label{fig:alter} 
\end{figure}

The binning design includes several pie charts to represent the proportion of participants(Figure~\ref{fig:alter}b). Pie areas in a pie chart are mapped to the numbers of data points in the grid. This design can clearly show the contribution of each participant, although it is not effective to visualize the global distribution.

The density map design is shown in Figure~\ref{fig:alter}c. Mayorga and Gleicher~\cite{splatterplots2013Mayorga} indicated that the bandwidth of kernel density estimation can be used as the abstraction metric.
Density maps can present the distribution intuitively.
However, when the number of participants increases, even the multi-class density map design~\cite{mcdensitymap2018Jo} is less effective to present distributions of multiple participants. Thus, we choose density maps to present the distribution.
It is overlaid with a lattice to support data selections by grids.

\textbf{Descriptive views.}
When regions are selected in the projection views, their attributes are presented in the descriptive views.
We use a bar chart (Figure~\ref{fig:interface}e) to present the contributions of each participant in the selected region. \jiazhi{This view helps participants to identify the sources of interested data, i.e., who owns the interested data.}
Parallel coordinates (Figure~\ref{fig:interface}f) are used to visualize the specific attributes of selected local points.
If the data type is image, the snapshots of local data replace parallel coordinates for data presentations.
Because of data privacy, participants can only visualize the high-dimensional representation of local data.

\section{Case Studies}
\label{sec:case}

In this section, we present three case studies to evaluate the effectiveness of SMAP.
The first two cases are conducted to simulate real-world distributed machine learning tasks.
We invited collaborators to perform in-lab studies, and observed and recorded the entire processes.
In the third case, SMAP was deployed to three hospitals where it was used and evaluated by real users in their daily work.

\subsection{Data Purchasing}
\jiazhi{The first case simulates a data market scenario where a data consumer (M) purchases data to improve his own small dataset to have enough data points in each class while avoiding introducing outliers. As being used in interactive labeling applications~\cite{Annotations2018Liu, bauerle2020classifier}, t-SNE aims to provide visual details, such as distribution of classes and outliers, which may be ignored by global statistics. Moreover, the joint embedding aligns the projections of consumer's data and provider's data and allows comparison between them.}

In this case, we used the Clothing 1M dataset~\cite{XiaoXiaetal2015} that contains 14 classes of cloth images (e.g., T-shirt, Shirt, Knitwear).
We assume that the data market finds three retailers (A, B, and C), each of which provides 3000 data points. The consumer M also has a smaller dataset of 1000 data points. A machine learning expert was invited to act as the consumer M in this case study.

\begin{figure}[!tb]
	\centering
    \includegraphics[width=0.7\columnwidth]{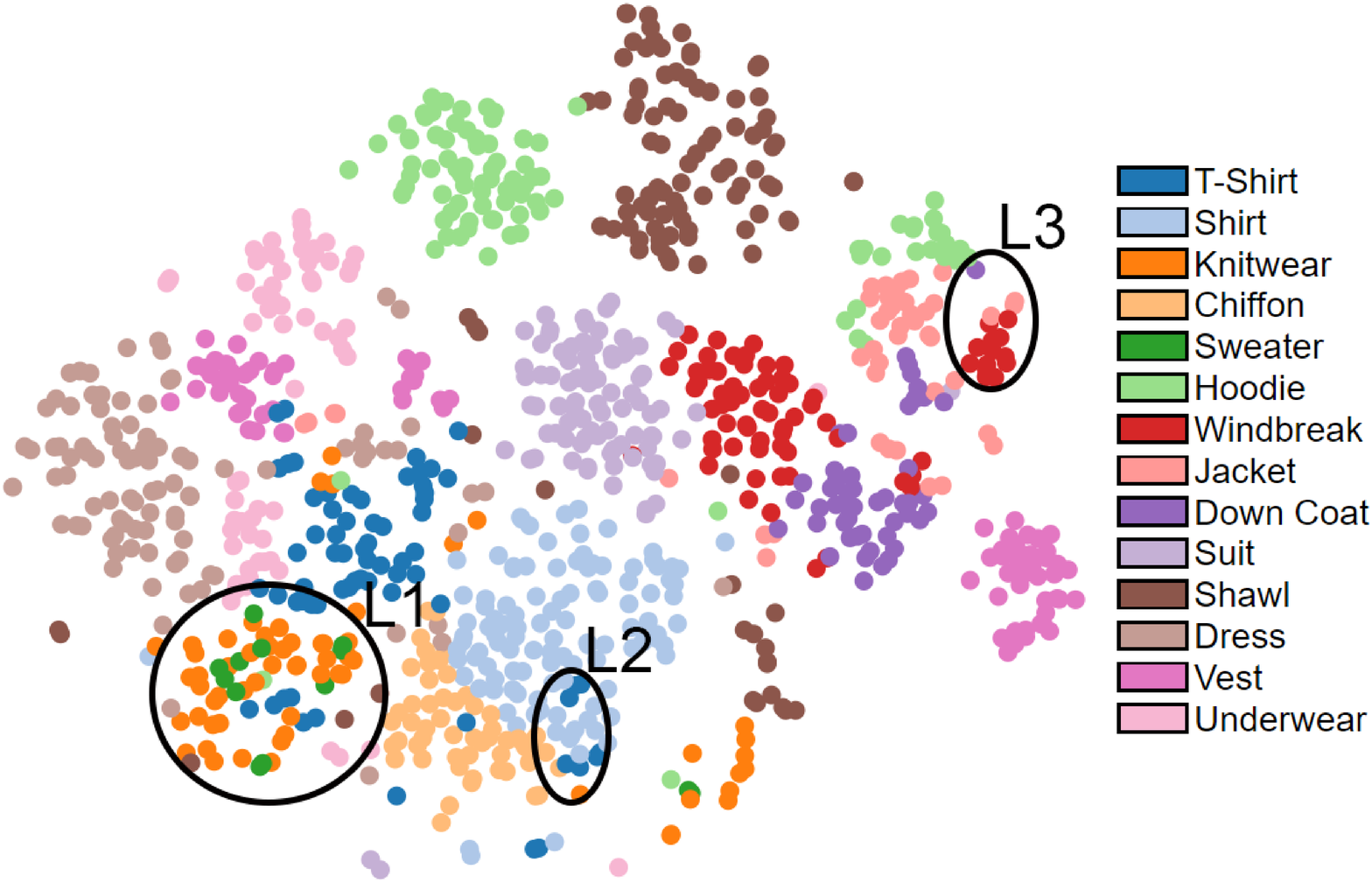}
	\caption{The t-SNE embedding of M's data.}
	\label{fig:case-market-user} 
\end{figure}

\jiazhi{After examining the local dataset with t-SNE (Figure~\ref{fig:case-market-user}), the expert found that the outliers (e.g., region L3 in Figure~\ref{fig:case-market-user}), the mixing of classes (e.g., region L1 in Figure~\ref{fig:case-market-user}), and the split of classes (e.g., region L1 in Figure~\ref{fig:case-market-user}) are the major issues.}

\begin{figure}[!tb]
	\centering
	\subcaptionbox*{(a1)\vspace{3pt}}{
	     \vspace{-3pt}
	     \hspace{-2mm}
	     \includegraphics[width=0.49\columnwidth]{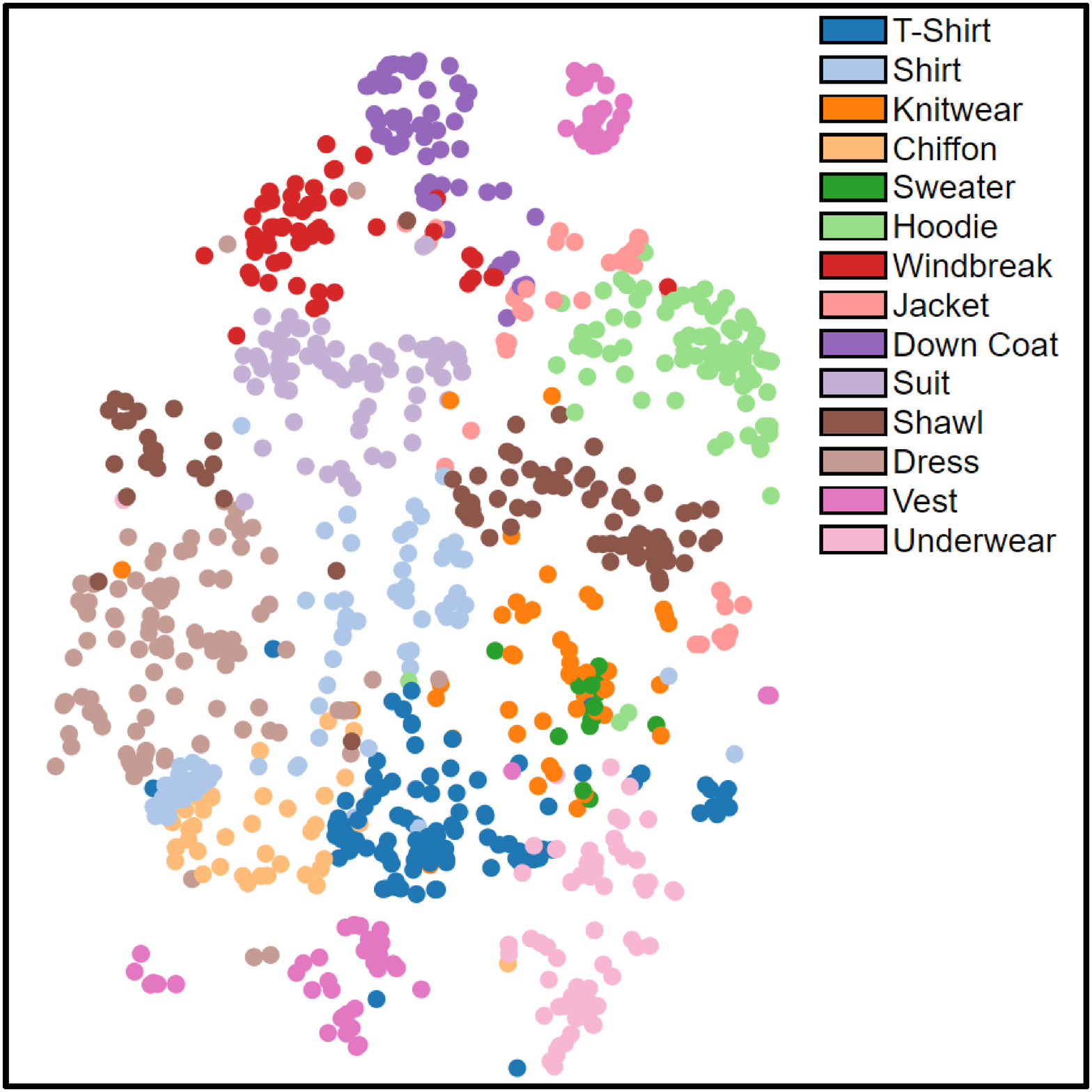}
	}
    \subcaptionbox*{(a2)\vspace{3pt}}{
	     \vspace{-3pt}
	     \hspace{-2mm}
         \includegraphics[width=0.49\columnwidth]{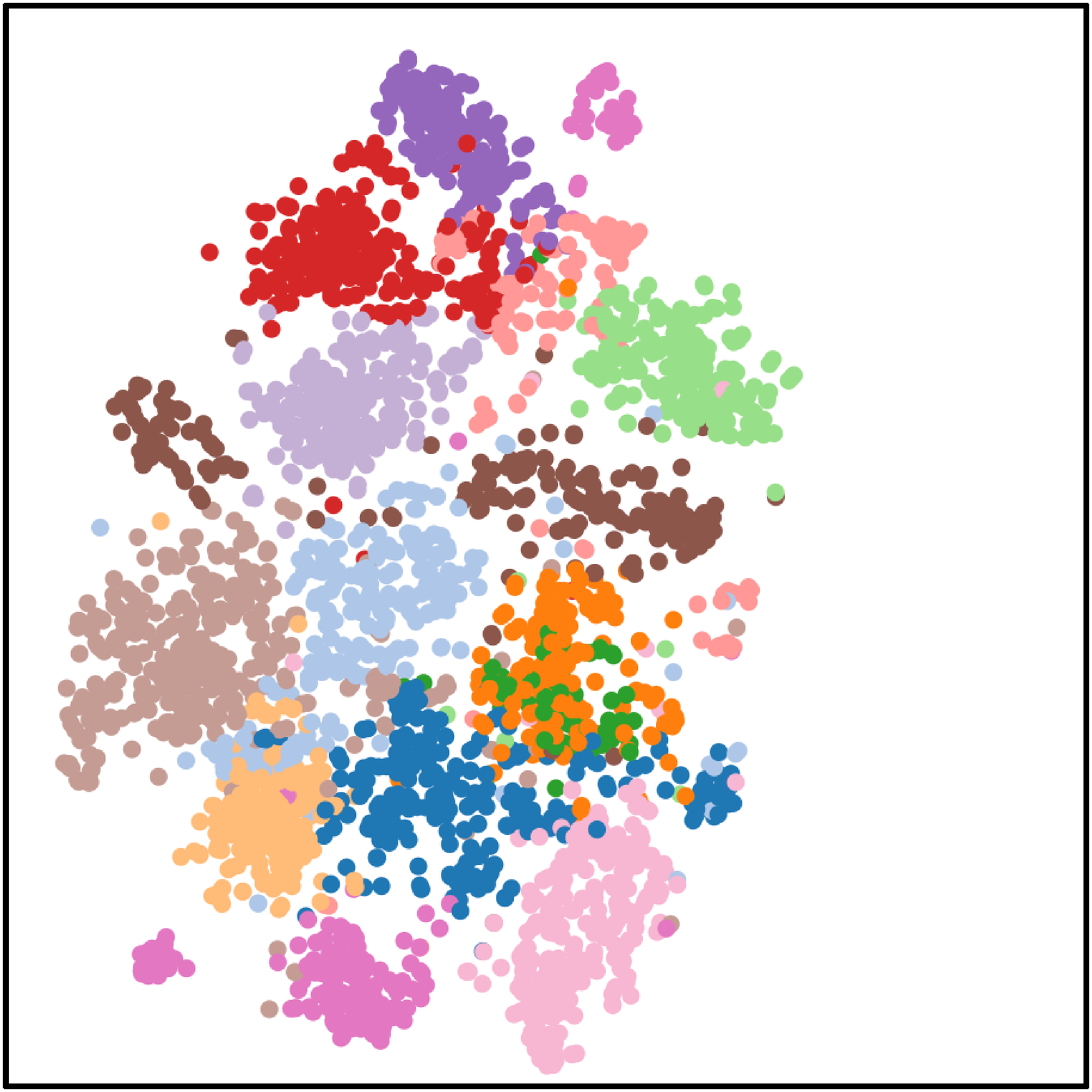}
	}
		\subcaptionbox*{\lei{(b1)}\vspace{3pt}}{
        \vspace{-3pt}
        \hspace{-2mm}
        \includegraphics[width=0.49\columnwidth]{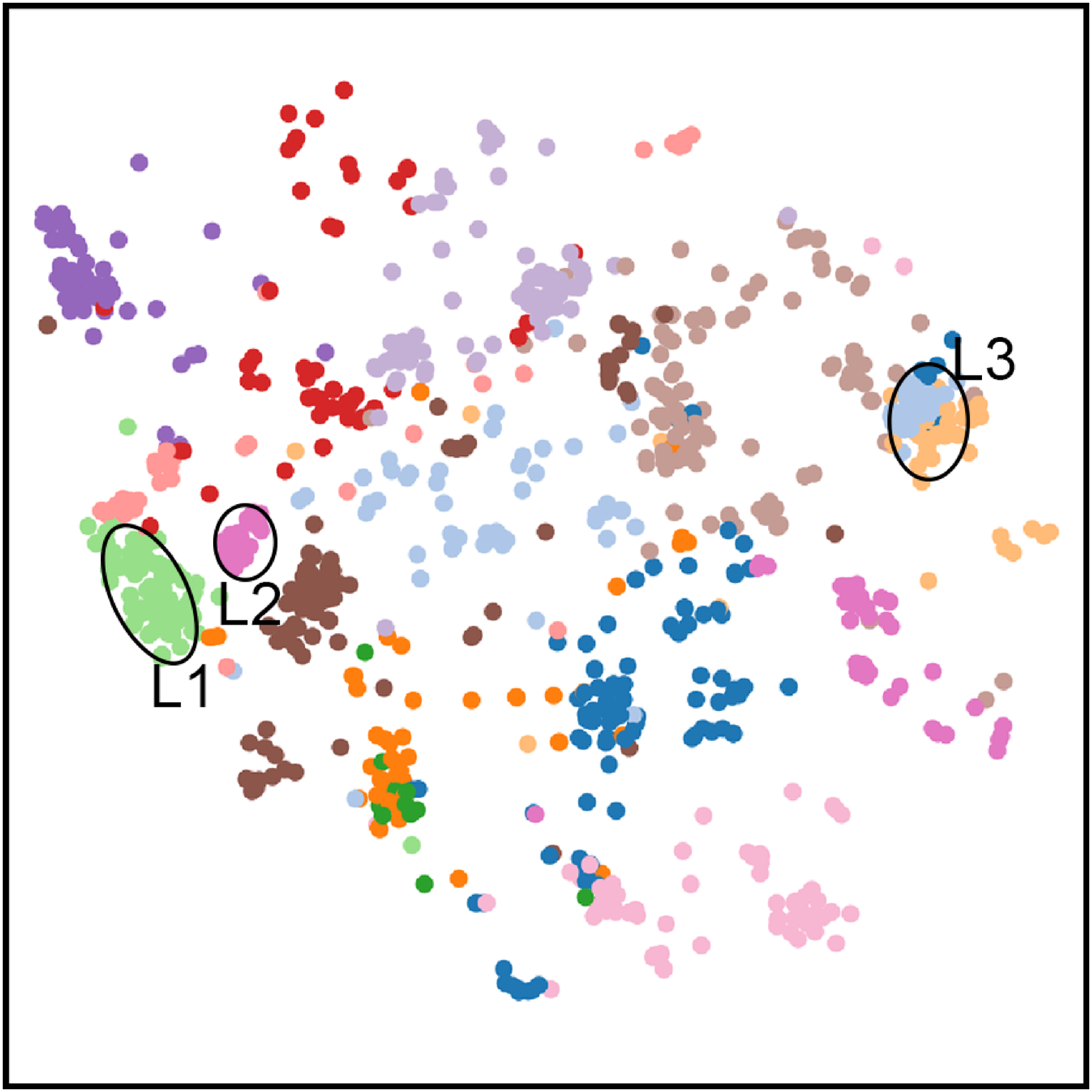}
	}
		\subcaptionbox*{\lei{(b2)}\vspace{3pt}}{
        \vspace{-3pt}
        \hspace{-2mm}
	    \includegraphics[width=0.49\columnwidth]{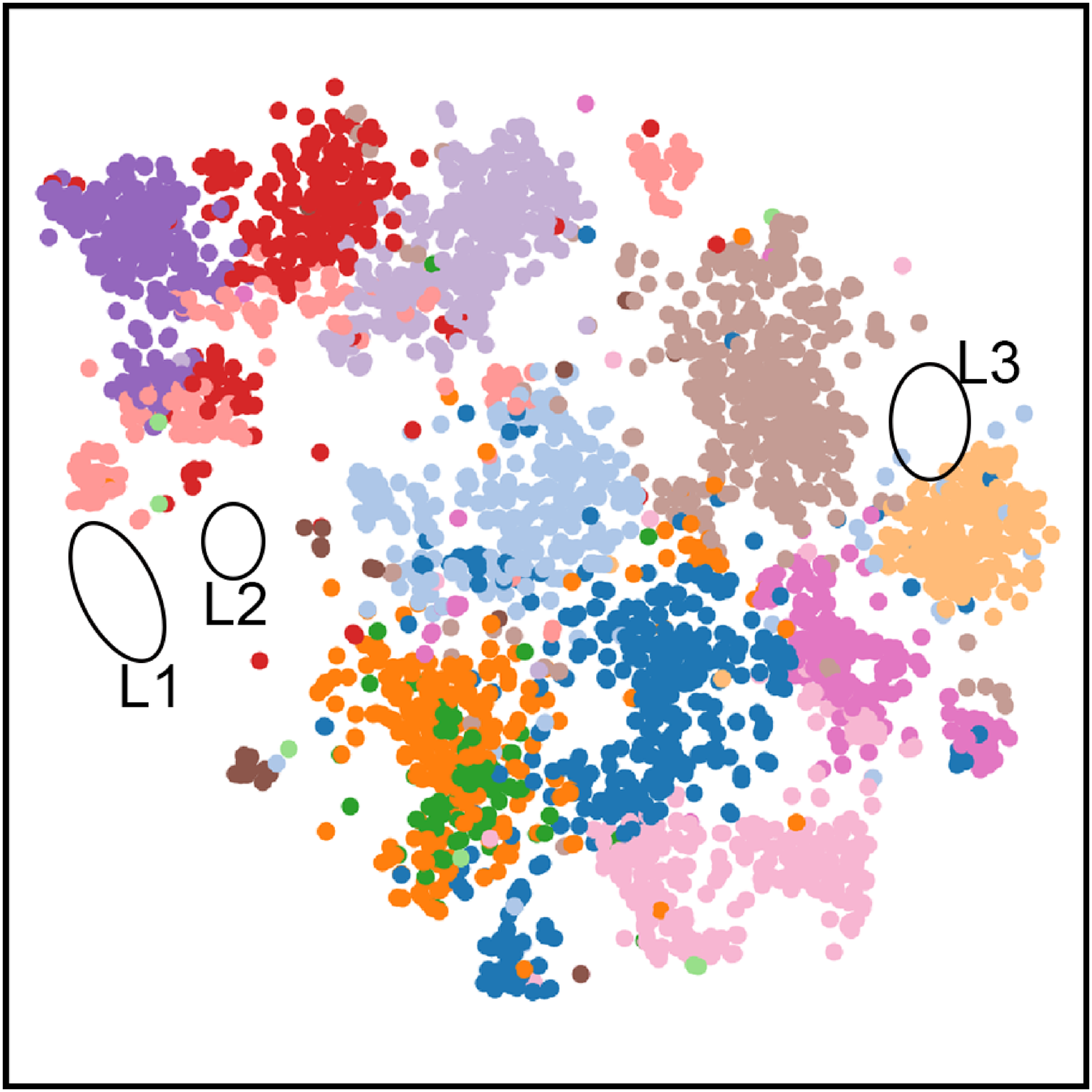}
	}
	
		\subcaptionbox*{(c1)\vspace{3pt}}{
        \vspace{-3pt}
        \hspace{-2mm}
        \includegraphics[width=0.49\columnwidth]{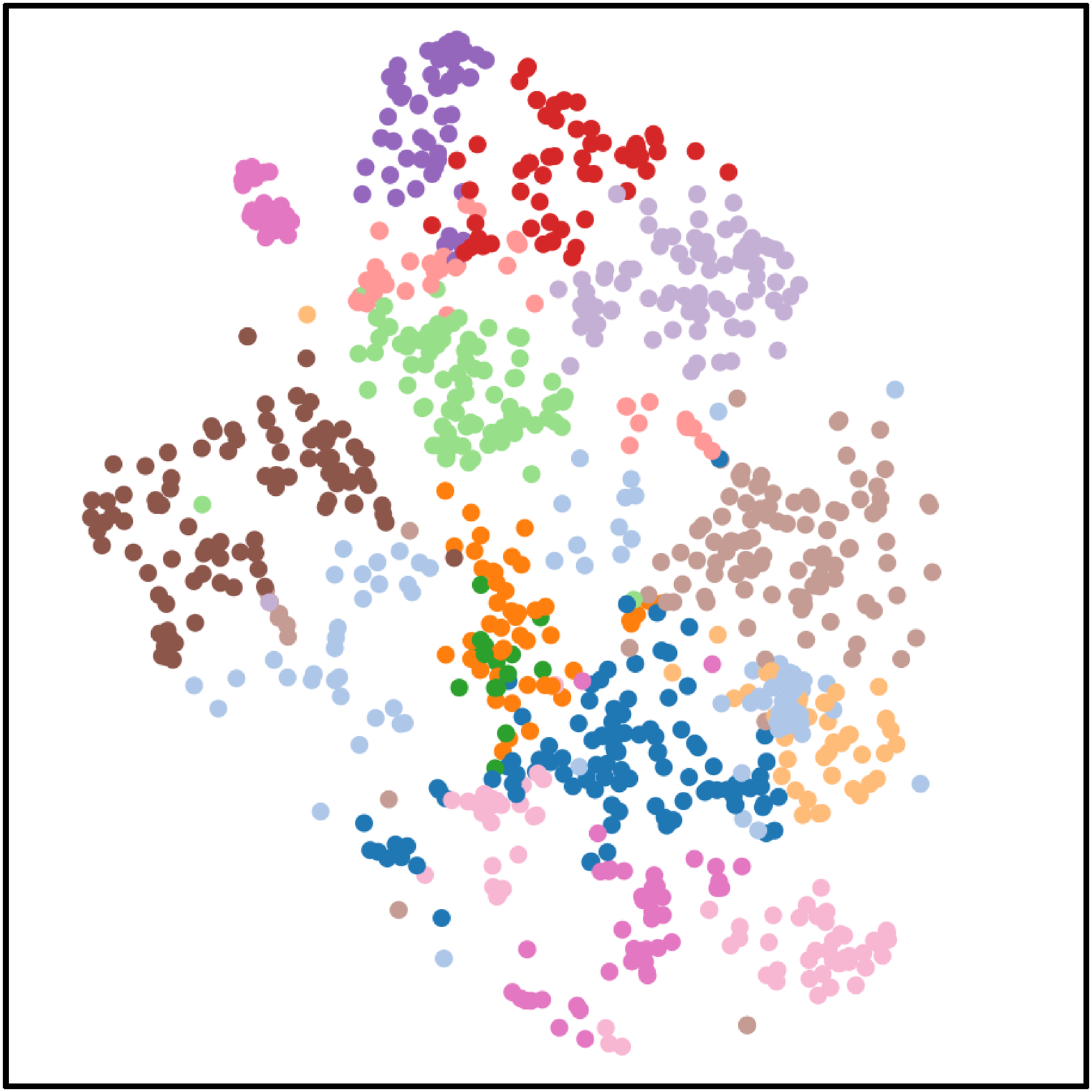}
	}
		\subcaptionbox*{(c2)\vspace{3pt}}{
        \vspace{-3pt}
        \hspace{-2mm}
	    \includegraphics[width=0.49\columnwidth]{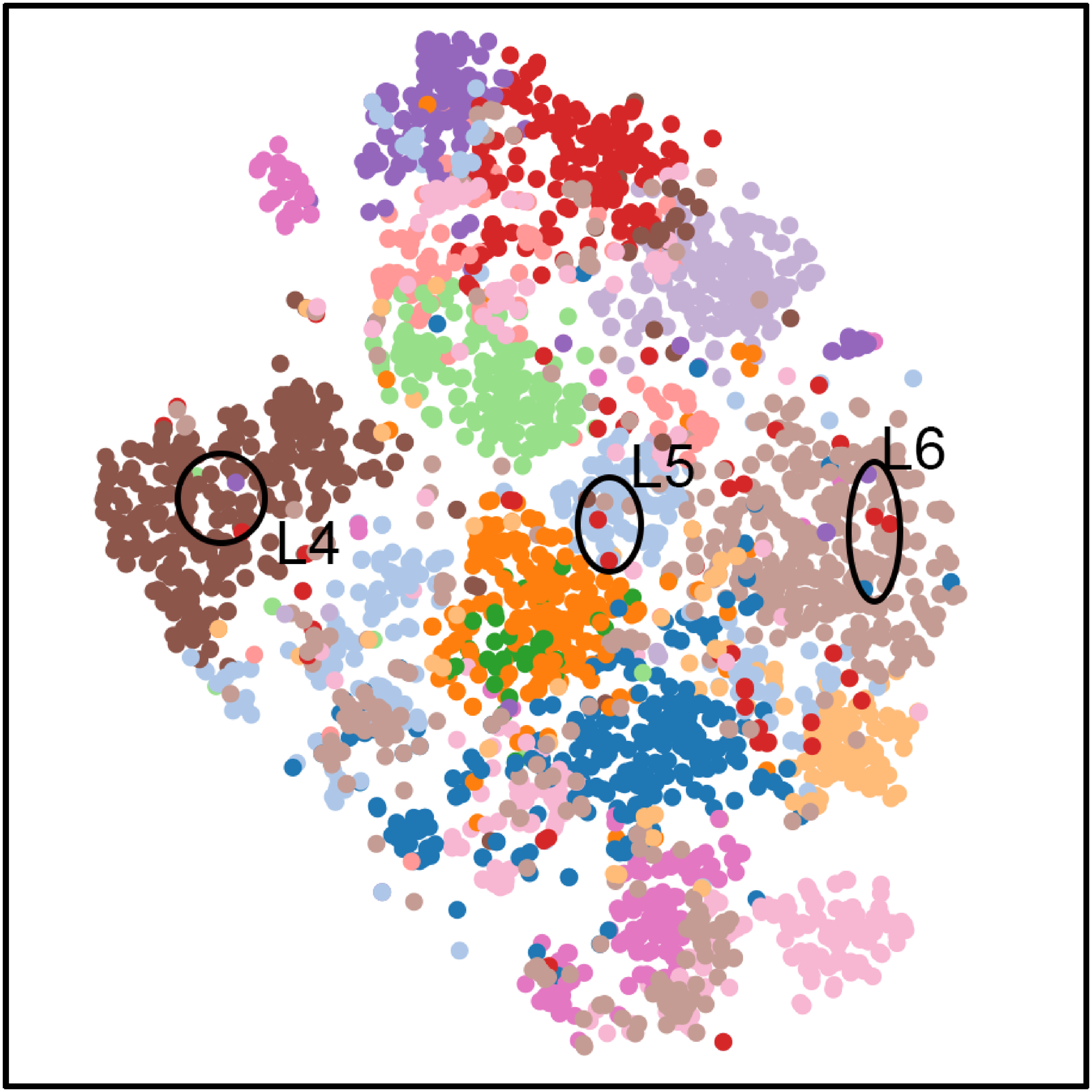}
	}
	\caption{The partial scatterplots of joint embedding results. The three rows are the results of joint embeddings between M and A (a1, a2), M and B (b1, b2), and M and C (c1, c2), respectively. The left column is the part of M's local data. The right column is the part of providers' data.}
    \vspace{-5mm}
	\label{fig:case-market} 
\end{figure}

Then, the expert was instructed to use the SMAP system for browsing and choosing appropriate datasets from the data market. In the task list, he focused on the three joint embedding tasks prepared by the providers of clothing datasets. They were displayed in scatterplots with class labels attached to the data points.
To explore the quality of these datasets, the expert iteratively joined in the three tasks for joint embeddings between his data and the providers' data datasets. Here, the embedding results were generated in advance due to the time-consuming computation.
Figure~\ref{fig:teaser} shows the joint embeddings of the three tasks.
\yang{The class structures (in terms of their labels and relative density) of the three datasets were generally the same as the expert's own dataset.}
Figure~\ref{fig:case-market} presents the partial scatterplots of the three joint embeddings where the expert observed some differences, especially for the dataset B. For example, the dataset B (Figure~\ref{fig:case-market}b2) lacked of some sub-cluster structures in comparison to Figure~\ref{fig:case-market}b1. 
This data insufficiency issue will result in performance degradation of the trained models~\cite{CoverSample20}.
The dataset C had all the important cluster structures (Figure~\ref{fig:case-market}c2) comparing with Figure~\ref{fig:case-market}c1.
However, there were many outliers (e.g., regions L4, L5, and L6 in Figure~\ref{fig:case-market}c2).
Based on the experience of examining M's local data, the expert made a hypothesis that there would be also many mislabeled data points in dataset C.
Figure~\ref{fig:teaser}a and Figure~\ref{fig:case-market}a2 
show that most local samples are covered by the dataset A appropriately.
In addition, the number of outliers in the dataset A is fewer than that in the dataset C.
As a result, he decided to purchase the dataset A.

\subsection{Multi-party Data Analysis for Identical Distribution}
\jiazhi{This case was conducted in a decentralized learning scenario inspired by Zhao et al.~\cite{zhao2018federated}. In this scenario, it is important that the data distribution of all participants are independent and identical distribution (IID). The non-IID issue will hinder the performance of further machine learning applications. The SMAP system is expected to be effective in handling this challenging task.}


\jiazhi{We used the MNIST dataset, which is also used in Zhao et al.~\cite{zhao2018federated}, and reduced its data dimension to 32 with PCA.
At the beginning, five different subsets containing 800 data points were distributed to five participants, respectively. All of them are master students with a research background in machine learning. They were instructed to use the SMAP to compare the data distribution with a secure joint embedding. To protect the point-level information of the data, the participants were asked to show the embedding results with density map.}




\begin{figure*}[!tb]
	\centering
    \includegraphics[width=2.0\columnwidth]{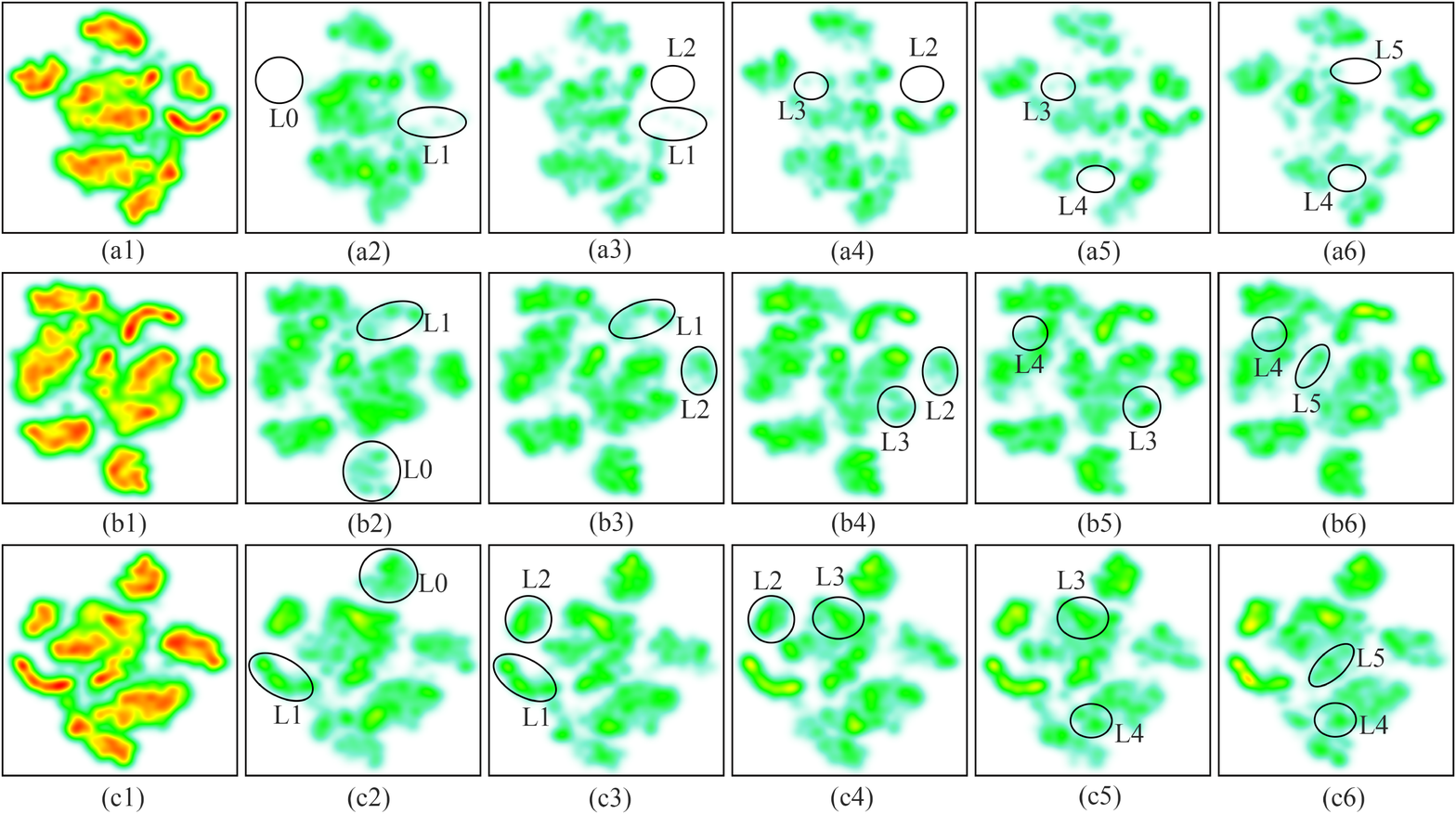}
	\caption{The results of joint embedding. (a1): the joint embedding result of five participants; (a2)--(a6): the partial density maps of participants A--E, respectively; (b1)--(b6): the joint embedding results of the uniform sharing strategy; and (c1)--(c6): the joint embedding results of the adaptive sharing strategy.}
	\label{fig:case-fl}
\end{figure*}

\jiazhi{Figure~\ref{fig:case-fl}a1 shows the joint embedding results for all data points.
The partial density maps of the participants A--E are shown in Figure~\ref{fig:case-fl}a2--a6, respectively.
When comparing the individual results with the global distribution, all participants found their data missed one or more cluster structures.  
The missing clusters are denoted by ellipses in Figure~\ref{fig:case-fl}.
For example, Figure~\ref{fig:case-fl}a2 shows that the data of Participant A is sparse in region L1 and absent in region L0.
From this view, Participant A concluded that two clusters were potentially missed in the data. After examining the results of other participants in SMAP, Participant A found that the local data in L2--L5 were missed in other participants' individual data. By performing the comparison analysis in SMAP, all participants were confident that the distributions of individual data were non-IID.}

\jiazhi{To ease the non-IID issue, the participants were introduced the data sharing strategy proposed in ~\cite{zhao2018federated}.
Specifically, each participant was asked to provide 120 points to compose a small shared subset of global data.  
To preserve data privacy, new data points were generated based on local distribution rather than exposing individual data directly.
To evaluate the efficiency of our approach in composing the shared subset, they 
compared two strategies for selecting local data.  
First, without using joint t-SNE, each participant was asked to perform a uniform sampling of their local data to create the shared subset. This strategy was proposed in ~\cite{zhao2018federated}.
The second strategy was guided by the joint t-SNE of SMAP. The participants adapted to share more data points in the region where other participants lacked data.
For example, Participant A could sample 120 points in region L2--L5 only rather than in the entire domain.
The two strategies were performed by the participants in two rounds of joint t-SNE. 
The results of the uniform strategy and adaptive strategy are shown in Figure~\ref{fig:case-fl}b1--b6 and Figure~\ref{fig:case-fl}c1--c6, respectively.
By visually comparing the heterogeneous regions of the results, all participants were confident that both strategies can ease the non-IID issue.
For example, the data of Participant A in region L0 increases in both Figure~\ref{fig:case-fl}b2 and Figure~\ref{fig:case-fl}c2.
Furthermore, in the results of the adaptive strategy (Figure~\ref{fig:case-fl}c2-c6), the distribution difference among participants is smaller than that of the uniform strategy (Figure~\ref{fig:case-fl}b2-b6).}

\subsection{Multi-party Data Analysis for Health Data}
In this case, we tested SMAP among three community hospitals.
These hospitals have health records of their patients from their annual health examination, and the records in each hospital represent the statistical characteristics of people from specific professions.
The community hospitals would like to analyze local data in the context of a more general population to promote the service for local communities.
However, due to the constraints of data privacy, they cannot fuse health data from multiple hospitals directly for joint analysis.

To address this issue, we provided SMAP to them for secure multi-party visual analysis.
We explained the mechanism of our secure multi-party visualization to them and signed contracts that ensure the security of local data.
To prevent the collusion between two collaborators, we serve as the collaborator T to hold encrypted data.
The collaborator S is served by an external company which is trusted by all hospitals.

\begin{figure}[!t]
	\centering
	\subcaptionbox*{(a)\vspace{3pt}}{
	     \vspace{-3pt}
	     \hspace{-2mm}
	     \includegraphics[width=0.49\columnwidth]{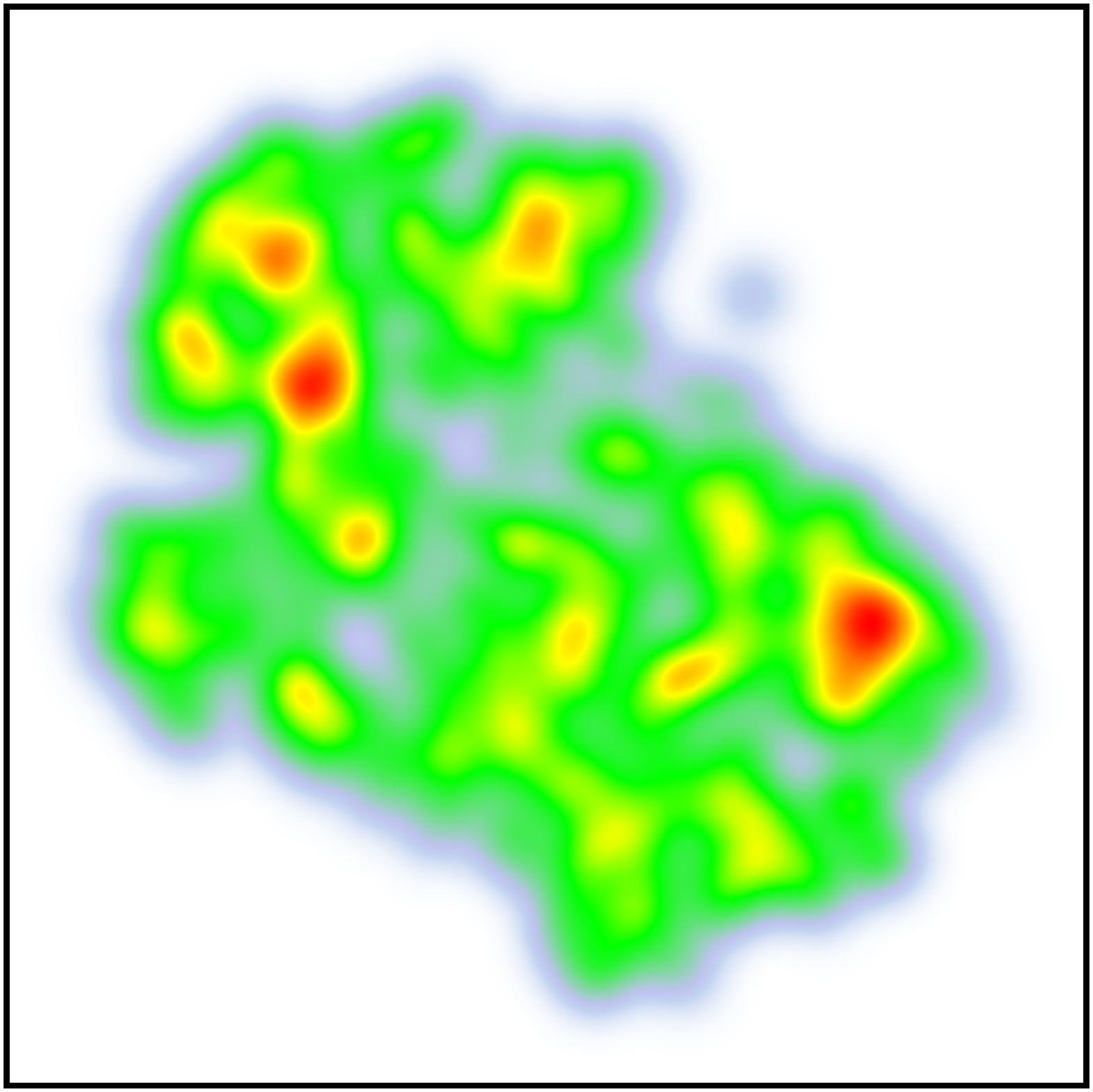}
	}
    \subcaptionbox*{(b)\vspace{3pt}}{
	     \vspace{-3pt}
	     \hspace{-2mm}
         \includegraphics[width=0.49\columnwidth]{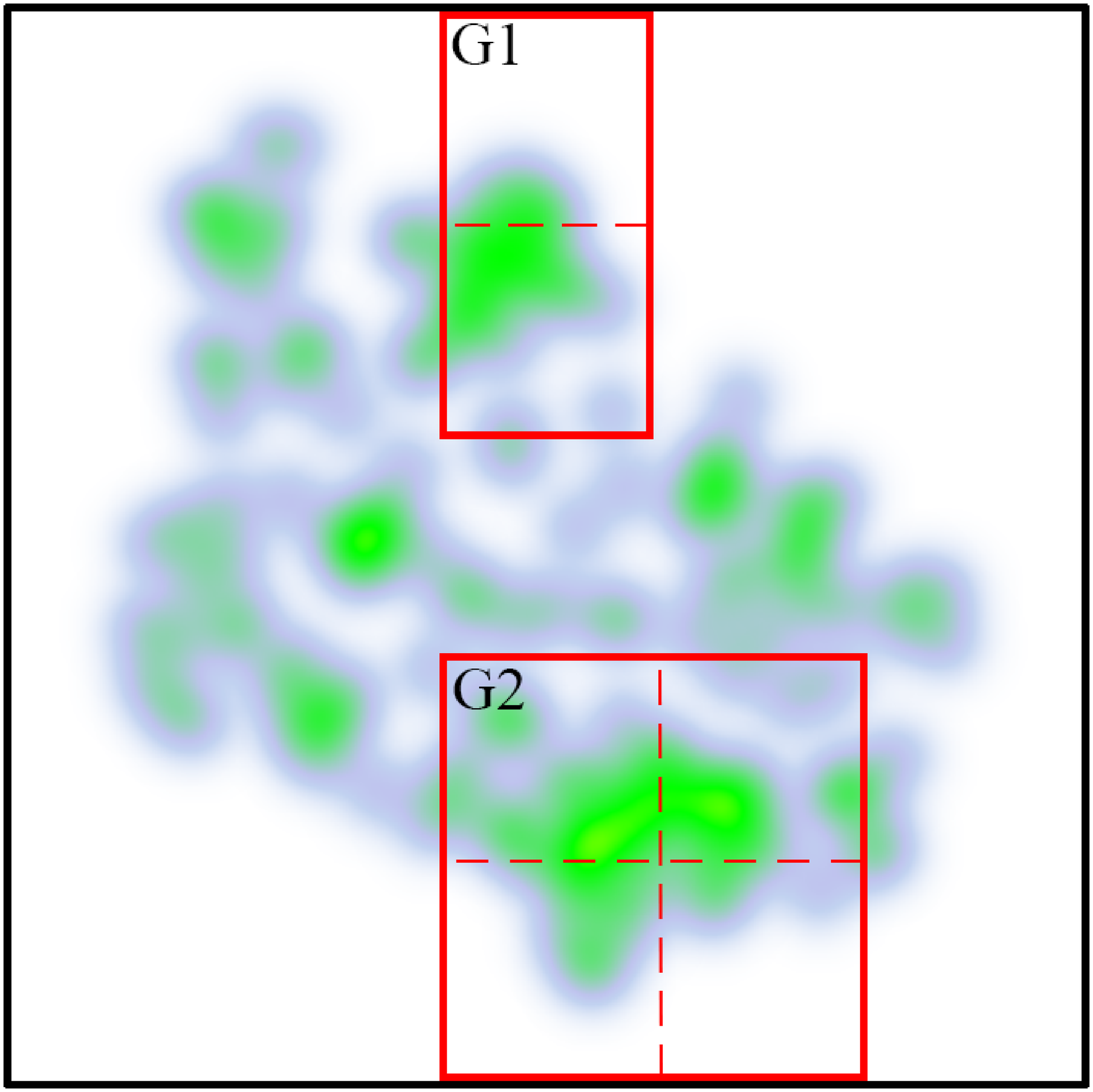}
	}
		\subcaptionbox*{(c)\vspace{3pt}}{
        \vspace{-3pt}
        \hspace{-2mm}
        \includegraphics[width=0.49\columnwidth]{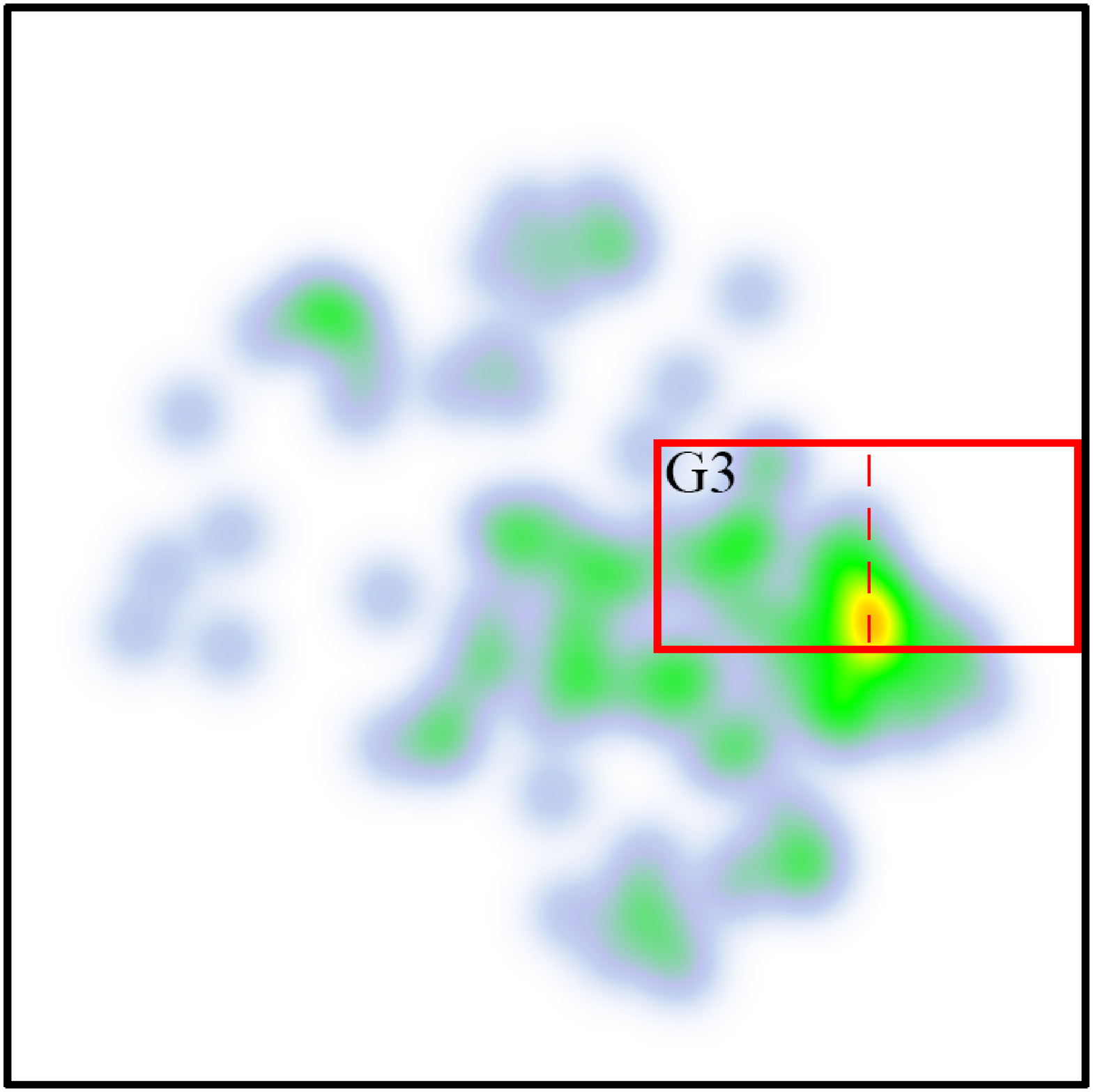}
	}
		\subcaptionbox*{(d)\vspace{3pt}}{
        \vspace{-3pt}
        \hspace{-2mm}
	    \includegraphics[width=0.49\columnwidth]{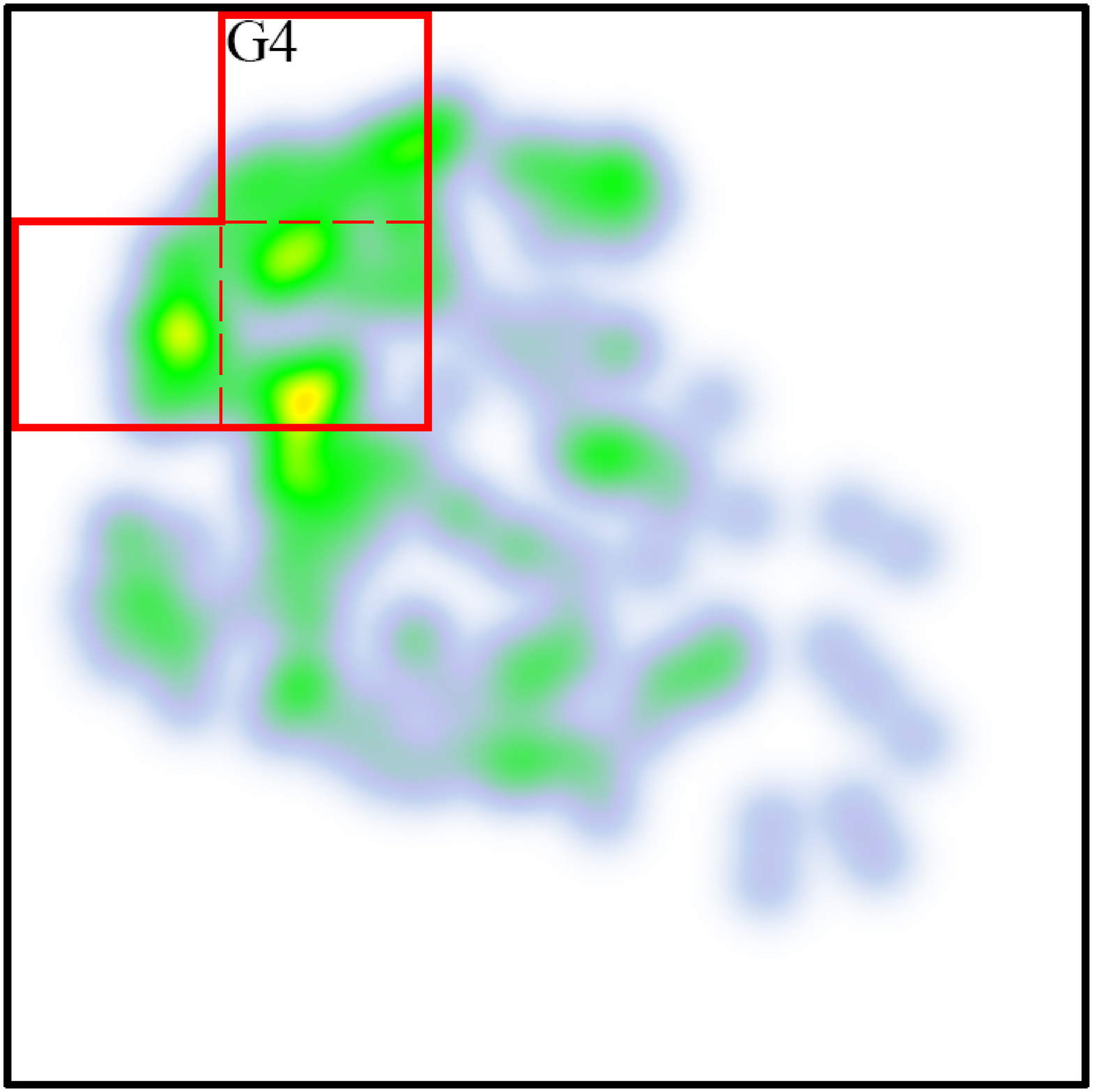}
	}
	\caption{The joint embedding results of health records. (a): the density map of all data; and (b)--(d): the partial density map of hospital A, B, and C, respectively.}
	\label{fig:case-health} 
\end{figure}

\begin{figure}[!t]
	\centering
	\subcaptionbox*{(a)\vspace{3pt}}{
	     \vspace{-3pt}
	     \hspace{-2mm}
	     \includegraphics[width=\columnwidth]{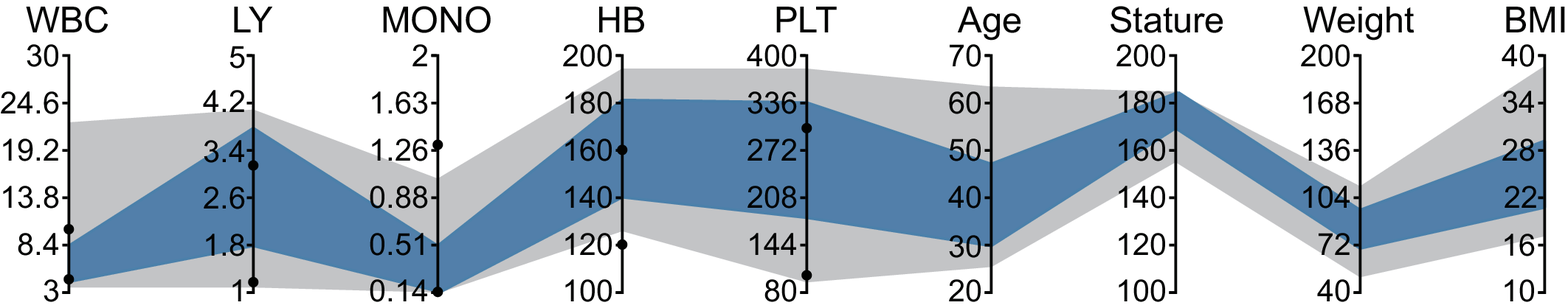}
	}
    \subcaptionbox*{(b)\vspace{3pt}}{
	     \vspace{-3pt}
	     \hspace{-2mm}
         \includegraphics[width=\columnwidth]{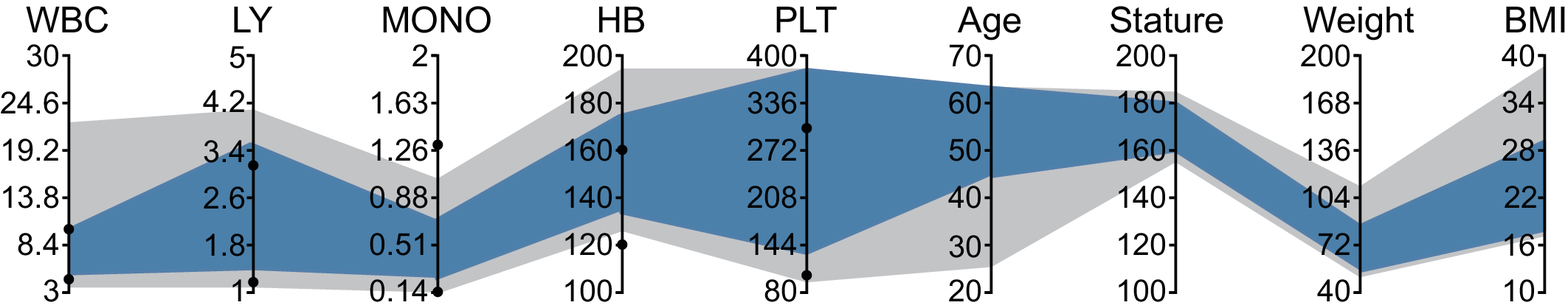}
	}
		\subcaptionbox*{(c)\vspace{3pt}}{
        \vspace{-3pt}
        \hspace{-2mm}
        \includegraphics[width=\columnwidth]{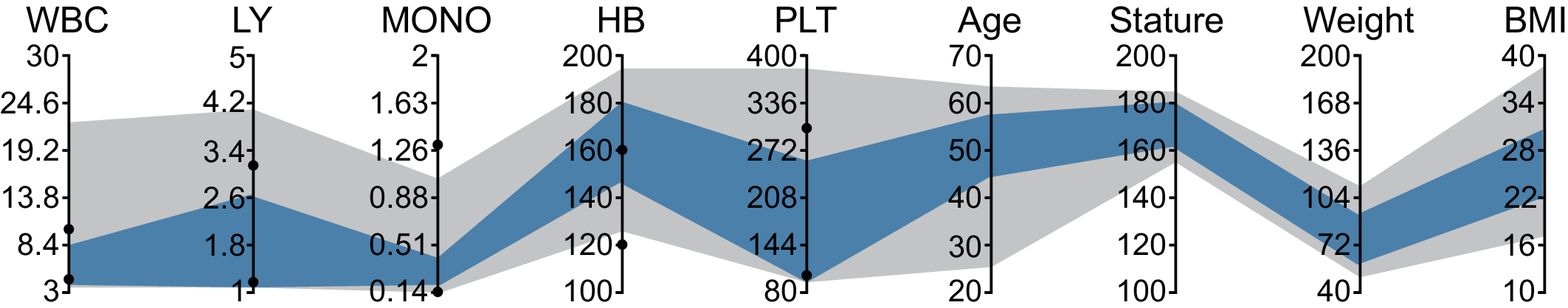}
	}
		\subcaptionbox*{(d)\vspace{3pt}}{
        \vspace{-3pt}
        \hspace{-2mm}
	    \includegraphics[width=\columnwidth]{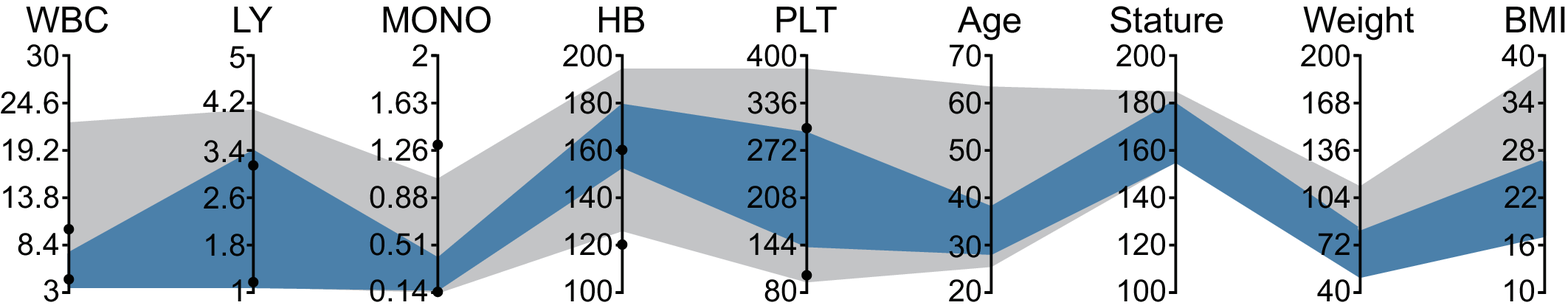}
	}
	\caption{Aggregated parallel coordinates of the selected local data. (a)--(d): local data in G1--G4, respectively. WBC: white blood cell count; LY: lymphocyte count; MONO: monocyte count; HGB: hemoglobin concentration; PLT: platelet count, BMI: body mass index. }
	\label{fig:case-health2} 
\end{figure}

Hospital A, B, and C sampled 
280, 107, and 159 records for male patients, respectively. These hospitals are the designated hospital by a design institution, a heavy industry factory, and a high-tech company, respectively.
The records have nine data dimensions, including the indices of physical examination and routine blood test.
\jiazhi{The data were normalized by dimension before the joint embedding. The three hospitals shared the range of each dimension to support the normalization.}
We collected the joint embedding results from the hospitals and requested hospital A to record their analysis process and snapshots.
The joint embedding results are presented in Figure~\ref{fig:case-health}.
The sub-maps of three institutions show that while three datasets covered similar areas, there are differences on the density distributions.
Hospital A was interested in these differences and explored them with the visual interface.
Four grided regions with salient differences are selected (G1--G4 in Figure~\ref{fig:case-health}b--d).
Note that although hospital A denoted the grids in the subgraph of other hospitals, the parallel coordinates show only local data.
Therefore, hospital A can only infer the distribution of other communities by local data.
\jiazhi{In additon, hospital A can know the contribution of each hospital in the interested region from the bar chart.}
The corresponding statistics are shown in the parallel coordinates (Figure~\ref{fig:case-health}a-d).
To preserve data privacy, hospital A sent us aggregated parallel coordinates rather than the original results.
The four grided regions represent four groups of people.
The distributions on age show that while people in the design institution are grouped into young designers (Figure~\ref{fig:case-health2}(a)) and senior designers (Figure~\ref{fig:case-health2}(b)), the factory has more young people (Figure~\ref{fig:case-health2}(d)) than the other two communities.
The Body Mass Index (BMI) indicates that people in the factory have lower BMIs.
It may be caused by more physically-demanding work in the factory.
A few records in G1, G2, and G4 exceed the upper limit of the normal value of the Lymphocyte Count (LY).
It may indicate chronic inflammation, which would be caused by long-time sitting of designers or physical labor by factory workers.
In contrast to people in G1, G2, and G4, people in G3 have a regular LY index.
With regard to PLT (platelet count), people in G1 and G2 have a higher index than other institutions.
It may indicate the risk of thrombus.
In conclusion, hospital A found that people in the high-tech company (mainly in G3) have more regular indices than the other two communities.
People in the design institution (G1 and G2) should be more active physically, while the factory community (G4) should rest more and have  balanced exercises.

\subsection{Expert Review}
\jiazhi{After each case study, we interviewed the experts and participants and summarized their comments. In case 1, the expert appreciated that the joint embedding provided aligned projections across data holders without having data from each other. The projections provided more information than statistics, such as the range, average, and variance of data, and made him more confident. He also commented that more type of projection algorithms, such as the PCA that he is familiar with, are expected to support more flexible analysis. In case 2, the five master students in machine learning appreciated the feature of the joint embedding in guiding the selection of shared data points. They agreed that it can enhance the flexibility and confidence of the data sharing strategy. In case 3, the expert in hospital A liked the design of our system. The learning cost is low because he is familiar with most plots , except the parallel coordinates graph. He indicated that it is flexible to compare distributions among three hospitals. He commented that it would be more promising if the hospital needs not to send encrypted data to the collaborator to fulfill stricter security rules. He pointed out that the computation speed was slow and he had to wait for 20 minutes for the embedding result.}

\section{Discussion}
\label{sec:discuss}

In this section, we discuss the characteristics of our approach, future research opportunities, and the limitation of our approach.

\vspace{3mm}
\noindent\textbf{Privacy-preserving embedding scheme.}
We identify two key issues, data decentralization and privacy preserving, as essential differences between the proposed joint embedding scheme and conventional embedding approaches. 
\jiazhi{Our contribution lies in the scheme for secure multi-party t-SNE. Specifically, our protocol is designed to compute the distance matrix across multiple participants while avoiding the exposure of the distance matrix to participants and collaborators. It would be interesting to apply our scheme to other dimensionality reduction approaches. Our scheme can directly support PCA, because it needs only to compute the covariance matrix rather than the distance matrix. The covariance matrix contains much less sensitive information. Applying our scheme to UMAP and ISOMAP may face some challenges, because for these two approaches, the distances among the k-nearest neighbors must be computed. These distances are a subset of the full distance matrix , and the fraction reduction cannot be performed in the step 7 of our protocol to protect the distance matrix. More research is need to evaluate the risk in disclose this subset.}



\vspace{3mm}
\noindent\textbf{Secure collaborative visual analysis.}
The proposed approach does not have a collaborative analysis implementation, because synchronous analysis and shared interactions are not required in our scenarios.
However, secure multi-party visualization provides a shared workspace.
All participants share the same visualization but have only the authority to access their own data.
A secure distributed collaboration can be naturally supported.
There are many scenarios of the new mode of collaborative visual analysis.
For example, the expert participated in our data market case suggested that voting or recommending for interesting data points would be useful in a crowdsourcing manner so that the consumers can estimate the price more accurately and confidently.
Without the secure multi-party visualization, however, it is impractical to explore the multiple attributes of data points for making effective recommendations.
Organizing secure collaborative visual analysis in different scenarios would be an interesting future work.

\vspace{3mm}
\noindent\textbf{Time performance.}
The proposed secure multi-party visualization scheme is time-consuming.
The most critical parts are data encryption/decryption and arithmetic operations on encrypted data.
The time complexity of computations on encrypted data is linear to the dimensionality \jiazhi{and the number of data points}.
In our implementation, in the case of ``Data Purchasing'', it takes about 32 hours to perform each task (4,000 points with 32 dimensions).
The 32 dimensional representation of Clothing 1M dataset is from Xiang et al.~\cite{correction19}.
In the case of ``Federated Learning'', the time for the task in round 1 (4,000 data points with 32 dimensions) is about 32 hours and about 50 hours for the task in round 2 (6,400 data points with 32 dimensions).
In the case of health data analysis, it takes 20 minutes to carry out the task (546 data points with 9 dimensions).

\jiazhi{Due to the high computation cost, we need to speed up the secure joint embedding. A practical solution is to employ high performance computing clusters. For example, we can find a typical set of cloud computing clusters that contain 300 3.2GHz virtual CPUs with 104 cores, which are 2971 times faster than the server in our implementation. If we equip each participant and collaborate with such a cluster and connect them with a 1.25GB/s network, the 
time cost for communications and computations could be reduced significantly. Specifically, Table~\ref{tab:time} presents the hypothetical time cost of each step in case 2 with such a cluster. The distance matrix related calculation in Step 3, 4, and 5 takes around 10s. With a speed of 1.25GB/s, the transition of encrypted noised distance matrix in step 4, which is the largest transferred data in the whole process, needs 14.5s. The whole protocol needs 29.2s for the transition.}

\begin{table}[!tb]
 \caption{\label{tab:time} The time cost in a hypothetical scenario with high performance computing clusters (300 3.2GHz CPUs with 104 cores).}
\begin{tabular}{c|cccccccc}
\toprule
 Step &  2 &  3 &  4 &  5 & 6 & 7 & 8\\
\midrule
  Time & 0.1s & 0.1s & 10.4s & 13.3s & 11.0s & 3.1s & 0.7s  \\
\bottomrule
\end{tabular}
\end{table}


\noindent\textbf{Overlaps in scatterplots.}
While overlaps in scatterplots is inevitable when number of points grows, the analysis of scatterplots-based embedding results, such as identifying outliers, is affected. In this paper, we rendered the data points in random order to ease this issue. In the future, we would like to try more solution. For example, we can render data points with transparency. We can also employ outlier detection algorthms and highlight the detected outliers.

\section{Conclusion and Future Work}
\label{sec:conclusion}

In this paper, we proposed a joint t-SNE scheme for secure multi-party visualization, implemented an online system for organizing and exploring joint embedding tasks, and tested the SMAP system through one case in real environment and two cases in laboratory environments.
Our major contribution is to develop a secure multi-party scheme for dimensionality reduction approaches.
Specifically, we proposed an instance of joint t-SNE.

There are several interesting future directions to address the limitations of our approach.
First, it is valuable to speed up secure multi-party projection.
The future breakthrough on secure multi-party could facilitate the advancement of secure multi-party projection.
Second, secure collaborative analysis enabled by joint embedding is also a promising direction.
Third, secure multi-party visualization for other data types is worth studying.

\acknowledgments{
We would like to thank the anonymous reviewers for their constructive suggestions.
This work was supported by National Natural Science Foundation of China (61872389, 61772456, 61761136020, 61972122) and Open Project Program of the State Key Lab of CAD\&CG (A1903).}

\bibliographystyle{abbrv-doi}

\bibliography{ref}
\end{document}